\documentclass[aip,jcp,reprint,longbibliography]{revtex4-2}
\usepackage{graphicx}
\usepackage{enumitem}
\usepackage{mathtools}
\usepackage{fixltx2e}
\usepackage{amsmath,amsthm,amsfonts,amssymb,amscd, fancyhdr, color, comment, graphicx, environ}
\usepackage{color}
\usepackage[normalem]{ulem}
\usepackage{caption}
\usepackage{subcaption}
\usepackage{comment}
\usepackage{soul}
\usepackage{bm}
\usepackage{array}
\usepackage{booktabs}
\usepackage{multirow}

\usepackage{amsmath}

\usepackage[breaklinks=true]{hyperref}
\usepackage{breakcites}

\usepackage{color, colortbl}
\usepackage{xr}

\definecolor{LightCyan}{rgb}{0.88,1,1}
\usepackage{colortbl}
\usepackage[table]{xcolor}
\usepackage{multirow} 
\definecolor{lightgray}{gray}{0.98}
\usepackage{makecell} 
\usepackage{tabularx}

\newcommand{\mfocus}[1]{\mathbf{#1}}
\newcommand{\Aib}{(Aib)\textsubscript{9}}
\newcommand{\methodname}{SPIBER}
\newcommand{\methodnamefull}{SPIBER (State Predictive Information Bottleneck-based Energy Reweighting)}

\begin{document}
	\title{\methodname: Reconstructing Free Energy Landscapes from Short, Unconverged Trajectories with Generative Flow Networks}

	\author{Venkata Sai Sreyas Adury}
	\affiliation{Chemical Physics Program and Institute for Physical Science and Technology,
		University of Maryland, College Park 20742, USA}
	
	\author{Pratyush Tiwary\footnote{Corresponding author.}}
	\email{ptiwary@umd.edu}
	\affiliation{Biophysics Program and Institute for Physical Science and Technology,
		University of Maryland, College Park 20742, USA}
	\affiliation{Department of Chemistry and Biochemistry and Institute for Physical Science and Technology,
		University of Maryland, College Park 20742, USA}
	\affiliation{University of Maryland Institute for Health Computing, Bethesda, Maryland 20852, USA}
	
	\date{\today}
	
	\begin{abstract}
		\textbf{Abstract: }
		Molecular systems have many degrees of freedom, but their metastable behavior can often be described by a few collective variables. Identifying these variables and estimating free energies along them from limited simulation data remains a challenging, important problem. Separate short trajectories may sample different metastable states without capturing transitions or establishing their relative equilibrium populations. For unbiased trajectories generated with the same Hamiltonian at a single temperature, alternate methods based on histogram reweighting cannot correct this imbalance. Here we present \methodname, which combines the State Predictive Information Bottleneck (SPIB) with Generative Flow Networks (GFlowNets). SPIB uses deep learning to approximate slow degrees of freedom through a past–future information bottleneck, retaining information needed to predict future metastable states. We show that this compression limits conditional entropy variations in populated regions, allowing conditional mean potential energies, which are much easier to calculate, to be used to approximate free energy differences. Given sufficient local sampling to estimate these energies, they define the target distribution for GFlowNets, energy-based generative samplers that sample according to estimated thermodynamic stability rather than observed populations. For a particle in a radial double-well potential, for alanine dipeptide, and for the nine-residue peptide \Aib, \methodname~recovers free energy differences between sampled metastable states to within one thermal energy unit of reference values. The method combines collective-variable learning and free energy estimation in up to four latent dimensions, without requiring converged state populations or additional molecular dynamics simulations.
	\end{abstract}
	
	\maketitle
	\renewcommand{\thesection}{\Roman{section}}
	\renewcommand{\thesubsection}{\Roman{section}.\Alph{subsection}}
	\renewcommand{\thesubsubsection}{\Roman{section}.\Alph{subsection}.\arabic{subsubsection}}
	\makeatletter
	\renewcommand{\p@subsection}{}
	\renewcommand{\p@subsubsection}{}
	\makeatother
	
	\section{Introduction}
	\label{sec:introduction}
	
	Molecular Dynamics (MD) simulations have become an indispensable tool for studying complex biomolecular systems, providing atomistic insight into processes that would be prohibitively difficult to reproduce experimentally. For an $N$-atom system, modern forcefields allow us to compute the potential energy $U(x)$ of any configuration $x \in \mathbb{R}^{3N}$ with relative ease. The challenge lies not in evaluating energies but in sampling: generating configurations according to their true Boltzmann probabilities, $p(x) \sim e^{-\beta U(x)}$, remains computationally intractable for most systems of practical interest.
	
	Fortunately, we rarely need the full high-dimensional distribution. Most biomolecular systems of interest spend the majority of their time in a small number of metastable states---conformations, binding modes, or functional states---separated by free energy barriers that make transitions between them rare events on accessible simulation timescales. This observation reduces the problem to two parts: finding a low-dimensional representation $\mfocus{z}: \mathbb{R}^{3N} \rightarrow \mathbb{R}^k$ ($k \ll 3N$), commonly called a \textit{collective variable} (CV), that separates the metastable states of interest, and estimating the equilibrium probability distribution $p(z)$ along this CV, or equivalently, the free energy surface (FES) $F(z) = -\frac{1}{\beta}\log p(z)$.
	
	The first part, finding CVs, has been addressed by numerous methods. When the states of interest are known \textit{a priori}, hand-chosen geometric or dihedral coordinates often suffice. More generally, linear methods like Principal Component Analysis (PCA)\cite{PCAOriginal}, Time-Structure Based Independent Analysis (tICA)\cite{PerezHernandez2013}, as well as many machine-learning-based methods\cite{Mehdi2024Enhanced} can distinguish metastable states from each other and even identify slow collective motions, given a simulated trajectory. These approaches work reasonably well for separating metastable states but do not guarantee that the resulting latent space is suitable for free energy estimation---a distinction we return to below.
	
	The second part, sampling the free energy surface, has a long history in statistical mechanics and computational chemistry. The most direct approach---running sufficiently long unbiased simulations until multiple transitions between states are observed---is rarely feasible when barriers are high. Enhanced sampling methods such as umbrella sampling\cite{torrie1977nonphysical} and metadynamics\cite{MetadynamicsOriginal} address this by applying external bias potentials along a chosen CV to accelerate exploration, followed by reweighting procedures to recover unbiased free energy estimates. While powerful, these methods require additional simulation time, careful parameter tuning, and \textit{a priori} knowledge of suitable CVs along which to apply the bias.
	
	When directly biased simulations are not desirable or feasible, reweighting methods like WHAM\cite{WHAM}, BAR, and MBAR\cite{BAR,MBAR} offer a theoretically rigorous framework for combining multiple trajectories into free energy estimates. In practice, however, they only provide meaningful corrections when running ensembles across a ladder of temperatures\cite{Wang_Berne_REST}, or by applying external biases such as umbrella potentials\cite{torrie1977nonphysical}. For collections of short, unbiased simulations performed at only one thermodynamic state condition, these methods have no advantage over simple histogramming. Recent work has attempted to circumvent the overlap and biasing requirement by constructing invertible maps from easily-sampled prior distributions to the target space\cite{InvPGM}, then applying MBAR in the transformed coordinates. While promising in principle, these approaches rely on the model learning an accurate global mapping---which is not guaranteed---and suffer from mode collapse and poor scalability to higher dimensions\cite{NormFlow_ModeCollapse}.
	
	In this work, we present a method, \methodnamefull,~that bypasses the need for additional molecular dynamics and biasing entirely by leveraging potential energy information directly along what we call \textit{entropy-free} CVs, rather than relying solely on population counts. The key insight comes from the thermodynamics-based decomposition of the free energy along a CV $\mfocus{z}$ as
	\begin{equation}
		\label{eqn:fundamental_relation}
		F(z) = \hat{E}(z) - T S(z),
	\end{equation}
	where $\hat{E}(z) = \int d^{3N}x\, U(x)\, p(x|\mfocus{z}=z)$ is the CV-conditioned average potential energy \cite{valsson2013thermodynamical}, readily estimated from simulation snapshots, and $S(z) = -k_B \int d^{3N}x\, p(x|\mfocus{z}=z)\log p(x|\mfocus{z}=z)$ is the conditional entropy. We show that when the CV is constructed using a well-regularized State Predictive Information Bottleneck (SPIB)\cite{SPIBOriginal}, the information-theoretic properties of SPIB ensure that $S(z)$ becomes approximately constant across $z$-space, while $\mfocus{z}$ stays quantifiably and mechanistically meaningful, reducing the free energy to:
	\begin{equation}
		\label{eqn:spiber_approx}
		F(z) \approx \hat{E}(z) + C,
	\end{equation}
	where $C$ is an additive constant. This means that potential energy alone---accessible from short, unconverged trajectories with no transitions between states---can provide meaningful estimates of free energy differences without any enhanced sampling.
	
	To sample from the resulting energy-based distribution in latent spaces of 3--5 dimensions (where traditional Monte Carlo becomes impractical), we employ Generative Flow Networks\cite{GFlowNetsOriginal, GFlowNetsFoundation} (GFlowNets) as policy-independent samplers. Unlike WHAM/MBAR, GFlowNets produce output distributions faithful to the reward function regardless of the input data distribution, making them ideal for reweighting non-overlapping samples\cite{EBGFN_paper_PMLR}.
	
	Even when all candidate metastable states have been visited briefly in multiple short MD simulations, obtaining a converged free energy surface is non-trivial owing to the difficulty in identification of a collective variable with sufficient overlap between separate trajectories, even assuming such a latent dimension exists.\\
	\methodname~addresses this issue by providing a systematic approach to construct the CV from first principles using only the existing MD data, while also maximally leveraging the information bottleneck to directly provide a solution to the reweighting problem in this CV space.
	
	We validate this approach on three benchmark systems of increasing complexity, demonstrating that SPIB-generated latent spaces combined with GFlowNet sampling provide accurate free energy estimates from short, unbiased trajectories alone. Notably, the method scales to four-dimensional latent spaces, where traditional energy-based sampling methods break down since they depend exponentially on the dimensionality\cite{MonteCarloReview}. GFlowNets scale only quadratically at inference time due to the principle of amortized inference\cite{GFlowNetsAmortization}. By construction, our protocol simultaneously optimizes both the CV, as shown here in up to 4 dimensions, and the quality of the free energy estimate---something not guaranteed by most existing pipelines where these two steps are decoupled. 
	
	\section{Methods}
	\label{sec:methods}
	Energy-based sampling methods (EBSMs) can generate samples from an unnormalized distribution $p(z) \propto e^{-\beta E(z)}$ for any energy function $E(z)$, provided the latent space $\mfocus{z}$ is sufficiently low-dimensional\cite{MetropolisMC, CBMC1}. In principle, one could apply such methods directly to physical systems by setting $E(z) = F(z)$, thereby sampling from the Boltzmann distribution $p(z) \sim e^{-\beta F(z)}$. This, however, is circular: it requires knowing the free energy before computing it. What we need instead is a computationally tractable approximation to $F(z)$. The decomposition in Eq.~\ref{eqn:fundamental_relation} splits the problem into two terms:
	\begin{equation}
		\label{eqn:energy_estimation}
		\hat{E}(z) := \int d^{3N}x\, U(x)\, p(x|\mfocus{z}=z)
	\end{equation}
	and the conditional entropy $S(z)$. The former is straightforward to estimate: $U(x)$ is available from any simulation snapshot, and as long as sampling at fixed $z$ is locally ergodic, the sample average converges to the true conditional mean. The latter term is far more challenging. We show below that information-theoretic arguments allow us to bound its variation across $z$-space---reducing it to an effectively constant shift---by minimizing the mutual information $I(\mfocus{X};\mfocus{z})$.
	
	\subsection{Methodically learning a CV for energy-based reweighting}
	\label{subsec:ideal_CV}
	A collective variable $\mfocus{z}: \mathbb{R}^{3N} \rightarrow \mathbb{R}^k$ (with $k \ll 3N$) maps the full configuration space to a low-dimensional latent representation. For our purposes, $\mfocus{z}$ must satisfy two requirements: (i) it should be mechanistically meaningful, i.e. be able to distinguish different metastable states as well as capture the correct sequence of state-to-state transitions, and (ii) it should yield a free energy estimate that is accurate up to an additive constant. We label the different metastable states by class variables $\mfocus{y}$. Strictly speaking, the approach in this work does not require the CV to capture the sequence of state-to-state transitions but we keep that as a desirable property for mechanistic interpretability of the CV. If we use an EBSM to sample from
	\begin{equation}
		\label{eqn:ebsm_energy_function}
		p(z) \sim e^{-\beta \hat{E}(z)} 
	\end{equation}
	with $\hat{E}(z)$ defined in Eq.~\ref{eqn:energy_estimation}, the resulting samples will reproduce the true Boltzmann distribution whenever the approximation of Eq.~\ref{eqn:spiber_approx} holds. We call CVs that satisfy this requirement entropy-free. We now derive the conditions under which a CV satisfies this requirement. First, we show how minimizing mutual information between input coordinates and the latent $\mfocus{z}$ can lead to entropy-free CVs. Then we show how the SPIB approach achieves this minimization while also satisfying the aforementioned condition of being mechanistically meaningful.
	
	\subsubsection{Minimizing Mutual Information yields Entropy-free CVs}
	\label{subsec:CVrules}
	The conditional entropy $S(\mfocus{z}=z)$ at any $z$ from Eq.~\ref{eqn:fundamental_relation} can be expressed in information-theoretic terms as
	\begin{equation}
		\label{eqn:entropy_estimation}
		\begin{split}
			\frac{S(\mfocus{z}=z)}{k_B} \equiv H(\mfocus{X}|\mfocus{z}=z) \\
			= -\int d^{3N}x\, \ln\left[p(x|\mfocus{z}=z)\right] p(x|\mfocus{z}=z)
		\end{split}
	\end{equation}
	which measures the uncertainty in the full configuration $\mfocus{X}$ once a value $z$ for $\mfocus{z}$ is chosen. For individual values $z$, the conditional entropy $H(\mfocus{X}\mid\mfocus{z}=z)$ may be either larger or smaller than $H(\mfocus{X})$. Its $p(\mfocus{z})$-weighted average equals
	\begin{equation}
		\label{eqn:conditional_entropy_mean}
		\mu \equiv H(\mfocus{X}) - I(\mfocus{X};\mfocus{z}) = \int d^{k}z\, p(z)\, H(\mfocus{X}|\mfocus{z}=z)
	\end{equation}
	where $I(\mfocus{X};\mfocus{z})$ is the mutual information between variables $\mfocus{X}$ and $\mfocus{z}$. As $I(\mfocus{X};\mfocus{z}) \to 0$, the weighted mean $\mu$ approaches the total entropy $H(\mfocus{X})$.\cite{InformationTheoryBasics}
	
	To quantify fluctuations of the conditional entropy over CV space, we discretize $\mfocus{z}$ into $M$ regions with probabilities $p(z_i)$ summing to unity, and define $\epsilon \equiv I(\mfocus{X};\mfocus{z})$. Let $h_i\equiv H(\mfocus{X}\mid\mfocus{z}=z_i)$, whose $p(\mfocus{z})$-weighted mean is $\mu$. The Bhatia--Davis inequality\cite{BhatiaDavis} states that a random variable $Y$ with mean $\langle Y\rangle$ and bounds $m\leq Y\leq M$ satisfies
	\[
	\operatorname{Var}[Y]\leq
	\bigl(M-\langle Y\rangle\bigr)
	\bigl(\langle Y\rangle-m\bigr).
	\]
	If $h_i$ were strictly confined to $[0,H(\mfocus{X})]$, this inequality
	and Eq.~\ref{eqn:conditional_entropy_mean} would immediately give
	\[
	\operatorname{Var}_{p(\mfocus{z})}[h]
	\leq I(\mfocus{X};\mfocus{z})
	\bigl[H(\mfocus{X})-I(\mfocus{X};\mfocus{z})\bigr],
	\]
	where we have used $M=H(\mfocus{X})$, $m=0$, and
	$\langle h\rangle_{p(\mfocus{z})}=\mu
	=H(\mfocus{X})-I(\mfocus{X};\mfocus{z})$
	
	The pointwise bounds $0\leq h_i\leq H(\mfocus{X})$ need not hold exactly in the most general case. $h_i$ is a Shannon entropy and therefore satisfies $h_i\geq0$ exactly. The upper bound $h_i\leq H(\mfocus{X})$, however, is not guaranteed pointwise: for an individual $z_i$, the conditional distribution can have greater entropy than the marginal distribution. Thus, only violations of the upper bound need to be controlled. However, we prove in the Appendix~\ref{appendix:proof} that as $I(\mfocus{X};\mfocus{z})\to0$, violations contribute only $o(1)$ to the mean and variance, and that for any region $\mathcal{R}$
	\begin{equation}
		\label{eqn:violation_limit}
		\sum_{i\in\mathcal R}p(z_i)\to0
		\qquad\text{whenever}\qquad
		\sum_{i\in\mathcal R}p(z_i)(h_i-\mu)^2 > 0.
	\end{equation}
	The Bhatia--Davis inequality therefore applies asymptotically, giving
	\begin{equation}
		\label{eqn:bhatia_davis_bound}
		\operatorname{Var}_{p(\mfocus{z})}[h]
		\leq I(\mfocus{X};\mfocus{z})
		\bigl[H(\mfocus{X})-I(\mfocus{X};\mfocus{z})\bigr]+o(1).
	\end{equation}
	Eq.~\ref{eqn:violation_limit} shows that driving $I(\mfocus{X};\mfocus{z})\to0$ forces
	the $p(\mfocus{z})$-weighted variance of $h_i$ to vanish. Since
	$\mu=H(\mfocus{X})-I(\mfocus{X};\mfocus{z})\to H(\mfocus{X})$,
	$h_i$ approaches $H(\mfocus{X})$, and therefore throughout regions of appreciable probability, $S(z)$ fluctuates very little over $\mfocus{z}$. Consequently, $S(\mfocus{z}=z_i)=k_Bh_i$ becomes effectively independent of $z_i$ over the populated CV space, allowing the entropic contribution to be absorbed into the constant $C$. The approximation $F(z)\approx\hat{E}(z)+C$ is therefore most accurate near the populated free-energy minima.
	The consequence of Eq.~\ref{eqn:violation_limit} is that the key argument underlying constant entropy CVs holds true except for regions where the population along the CV is exceedingly small. High-population regions with $p(z_i)$ non-vanishing, as quantified in the sense of Eq.~\ref{eqn:violation_limit}, exhibit the smallest deviations from the mean entropy. Metastable states correspond precisely to these populated regions (free energy minima), where the approximation $S(z) \approx \mathrm{const}$ is most accurate.\\
	Empirically, we find this approximation works well even for regions corresponding to barriers in the free energy surface, as seen from the results in Sec.~\ref{sec:results}.\\
	Even though these results are derived for an arbitrary partitioning $\mfocus{z}$, reweighting in $\mfocus{z}$ transfers directly to the metastable states themselves, provided $\mfocus{z}$ groups the relevant metastable states into contiguous regions.
	
	\subsubsection{SPIB gives entropy-free and mechanistically meaningful CVs}
	So far, we have shown that reducing the mutual information between the high-dimensional feature space $\mfocus{X}$ and low-dimensional CV $\mfocus{z}$ leads to minimizing the location-dependence of the entropy $S(z)$. A trivial way of doing so would be to set $\mfocus{z}$ to be an $\mfocus{X}$-independent Gaussian random variable. In such a case, $\mfocus{z}$ carries no predictive power for the underlying chemistry or physics of the system. Thus our challenge here is two-fold: we seek a CV $\mfocus{z}$ with constant entropy that still provides mechanistic insight into the system being studied. The central idea in SPIB is metastability. SPIB assumes that a system transitions between metastable states, with a separation of timescales between relaxation within each state and transitions between states. After entering a metastable state, the system loses memory of its initial configuration on a timescale much shorter than its typical residence time in that state. The number of metastable states depends on the temporal resolution, $\Delta t$, at which the dynamics is studied. At small $\Delta t$, more states can be resolved. As $\Delta t$ increases, faster motions are averaged out, reducing the number of distinguishable metastable states. SPIB provides a principled way to construct such a collective variable $\mfocus{z}$ from short, pre-existing MD trajectories that provides both of these properties. SPIB's goal is to extract the minimum amount of information needed to predict the final metastable state label $\mfocus{y}$ after time $\Delta t$, where $\mfocus{y}$ is a one-hot encoded\cite{onehot_encoding} state vector. This is mathematically represented as the objective function for SPIB\cite{SPIBOriginal,ShamsSPIB}:
	\begin{equation}
		\label{eqn:SPIB_loss}
		\mathcal{L}_{IB} \equiv I(\mfocus{z};\mfocus{y}) - \beta_{IB}\, I(\mfocus{X};\mfocus{z}),
	\end{equation}
	which balances two competing goals. Maximizing $I(\mfocus{z};\mfocus{y})$ ensures that $\mfocus{z}$ retains enough information to distinguish the metastable states (labeled by $\mfocus{y}$), while minimizing $I(\mfocus{X};\mfocus{z})$ compresses away everything else.  As shown in Sec.~\ref{subsec:CVrules}, this also makes the entropy $z$-independent.\\
	
	The predictive part of SPIB's loss function in Eq.~\ref{eqn:SPIB_loss} is $I(\mfocus{y};\mfocus{z})$, which deals with state-classification, and makes $\mfocus{z}$ actually a useful CV. The regularization term, whose strength is controlled by $\beta_{IB}$, caps $I(\mfocus{X};\mfocus{z})$, making our energy-based reweighting possible. Together, these trade-offs allow us to achieve rapid estimation of free energy differences between metastable regions. We refer to our previous work\cite{SPIBOriginal} and Appendix \ref{appendix:SPIB} for details of the SPIB approach.
	
	When regularization is sufficiently tight that $I(\mfocus{X};\mfocus{z}) < \epsilon$ for some small $\epsilon$, Eq.~\ref{eqn:bhatia_davis_bound} gives:
	\begin{equation}
		\label{eqn:SPIB_to_BhatiaDavis}
		Var[H(\mfocus{X}|\mfocus{z})] \leq \epsilon\,\big(H(\mfocus{X}) - \epsilon\big),
	\end{equation}
	which vanishes asymptotically as $\epsilon \rightarrow 0$. This establishes why a well-regularized SPIB produces CVs suitable for energy-based free energy estimation: the entropy correction becomes negligible, leaving $F(z) \approx \hat{E}(z) + C$.
	
	\subsubsection{Other dimensionality reduction methods}
	We emphasize here that SPIB, as seen through Eq.~\ref{eqn:SPIB_loss}, naturally performs the task we set out to do - it minimizes mutual information between input features and the CV while ensuring that the CV stays mechanistically meaningful. It is still natural to wonder if this dual objective could be achieved through other dimensionality reduction methods.  Contrasting SPIB with other widely used dimensionality reduction techniques, Principal Component Analysis (PCA) and Variational Autoencoders (VAEs) share an opposite, common philosophy: preserve \textit{as much information as possible} about $\mfocus{X}$ in the latent representation. PCA achieves this linearly by finding orthogonal directions of maximal variance; under Gaussian assumptions, it is equivalent to maximizing $I(\mfocus{X};\mfocus{z})$ exactly. VAEs generalize the compression analogy to non-linear encoders by optimizing the evidence lower bound (ELBO)~\cite{KingmaWelling2013}. 
	
	Performance of a VAE is shown to be linked to maximizing --- as opposed to minimizing --- $I(\mfocus{X};\mfocus{z})$\cite{DVIB,InfoVAE}. This is the opposite of what SPIB does: SPIB \textit{minimizes} $I(\mfocus{X};\mfocus{z})$ to suppress entropy variation, while VAEs and PCA \textit{maximize} it to preserve reconstruction fidelity. In practice, however, strategically discarding information is a far more tractable objective than preserving all of it. Methods that maximize $I(\mfocus{X};\mfocus{z})$ typically require many more latent dimensions to capture metastable structure and are generally unsuitable even for most EBSMs.
	
	A third popular approach is the time-lagged Independent Component Analysis (tICA)~\cite{NaritomiFuchigami2011, PerezHernandez2013}, which pursues yet another objective: finding linear combinations of features that maximize autocorrelation at a chosen lag time, thereby capturing slow collective motions. Unlike with PCA and VAE, here the effect on mutual information $I(\mfocus{X};\mfocus{z})$ is not obvious.
	
	We benchmark all three alternatives --- PCA, VAE, and tICA --- against SPIB for the same latent variable dimensionality $k$ across our test systems (see Sec.~\ref{subsec:results-other-methods}). As predicted by theory in Sec.~\ref{subsec:CVrules}, SPIB strictly outperforms them for free energy estimation. PCA performs poorly in practical cases, while tICA behaves similarly to an unregularized SPIB: it identifies sharp state boundaries but produces projections unsuitable for reweighting.
	
	\subsection{Estimating $\hat{E}(z)$}
	\label{subsec:energy_estimation}
	Formally, the CV-conditioned average potential energy is a Boltzmann-weighted average over all configurations mapping to a given $z$:
	\begin{equation}
		\label{eqn:energy_estimation_meaning}
		\hat{E}(z) := \frac{\sum_{x\in\mathbb{R}^{3N}} e^{-\beta E(x)}\, E(x)\, \mathbb{I}[\mfocus{z}(x) = z]}{\sum_{x\in\mathbb{R}^{3N}} e^{-\beta E(x)}\, \mathbb{I}[\mfocus{z}(x) = z]},
	\end{equation}
	where $\mathbb{I}[\cdot]$ is the indicator function that selects only configurations with latent coordinate $z$. In practice, we approximate this using a kernel-based estimator: for any query point $z^*$ in latent space, we collect all simulation frames whose encoded coordinates fall within a radius $R_0$ and average their potential energies:
	\begin{equation}
		\label{eqn:energy_estimation_practical}
		\hat{E}(z^*) \approx \mathbb{E}_{x \sim \text{data}}\left[ E(x) \;\big|\; |\mfocus{z}(x) - z^*|^2 \leq R_0^2 \right].
	\end{equation}
	The choice of $R_0$ and the minimum sample threshold required for a valid estimate are discussed in the protocol section below (Sec.~\ref{subsec:protocol}).
	
	\subsection{Energy-based Sampling with GFlowNets}
	\label{subsec:gfnet_sampling}
	Now that we have established that SPIB can learn a CV that has minimum position-dependence of associated entropy while still being predictive of future metastable states, we still need a method to sample along this CV. Given an energy function $E(z)$ over a low-dimensional latent space, sampling from the Boltzmann distribution $p(z) \propto \exp[-\beta E(z)]$ is straightforward in principle. The most direct approach would be to run Langevin dynamics in latent space~\cite{LangevinDynamics}, but this inherits the same barrier-crossing limitations as physical simulations --- even a low-dimensional CV can exhibit free energy barriers too high for equilibration within reasonable time. Rejection sampling with Monte Carlo~\cite{RejectionMC} avoids dynamics altogether, but its acceptance rate decays exponentially with dimension since the latent space becomes sparse. Markov-chain Monte Carlo~\cite{MetropolisMC} mitigates this by focusing on populated regions, yet still struggles to jump between well-separated metastable basins when no prior knowledge of their relative positions is available. We need a sampler that can leverage existing trajectory data to focus on relevant regions while retaining the global coverage of rejection sampling. This niche is met by Generative Flow Networks (GFlowNets).
	
	GFlowNets are generative models trained to sample objects from an unnormalized target distribution~\cite{GFlowNetsFoundation, GFlowNetsOriginal}. In the reinforcement-learning formulation, each object is constructed step-by-step, and a reward is assigned only upon completion. By setting this reward to $R(z) = \exp[-\beta \hat{E}(z)]$, the GFlowNet learns to sample latent configurations proportional to the Boltzmann distribution, effectively acting as an energy-based sampler~\cite{EBGFN_paper_PMLR}. For low-dimensional CV spaces (1--2D), traditional Markov-chain Monte Carlo methods can achieve equivalent sampling, but their convergence time grows exponentially with dimension. GFlowNets circumvent this by amortizing the sampling cost into a learned policy, making free-energy estimation tractable in 3--5 dimensional latent spaces where classical enhanced-sampling techniques typically struggle~\cite{GFlowNetsFoundation, GFNetHighD}.\\
	
	A critical advantage of energy-based GFlowNets is population independence: the output distribution depends only on the reward function, not on the distribution of the training data. For an entropy-free CV, $F(z) \approx \hat{E}(z)+C$. $\hat{E}(z)$ is tractable from MD snapshots, calculated as described in Eq.~\ref{eqn:energy_estimation} using samples from the neighborhood of the chosen point $z$. Thus, this $\hat{E}(z)$ can be chosen as the reward for generating any coordinate $z$ which will allow the GFlowNet to reproduce the Boltzmann distribution.
	
	This lets us reweight multiple short, unbiased trajectories with incorrect starting populations and still recover the correct thermodynamic distribution, for all reasonably populated regions of $\mfocus{z}$ where enough samples exist to estimate $\hat{E}(z)$. This contrasts sharply with standard reweighting methods (WHAM, MBAR), which require substantial overlap between sampled regions of different states, and with diffusion models, which faithfully reproduce their training distribution but are difficult to steer toward a target distribution\cite{DiffusionScoreModels}.  The only requirement is local ergodicity within each metastable state: if the CV separates the key states and the neighborhood around each $z$ is adequately sampled, then $\hat{E}(z)$ can be reliably estimated from nearby configurations.
	
	\subsection{Complete \methodname~protocol}
	\label{subsec:protocol}
\begin{figure*}
		\captionsetup{skip=6pt}
		\centering
		\includegraphics[width=0.95\textwidth]{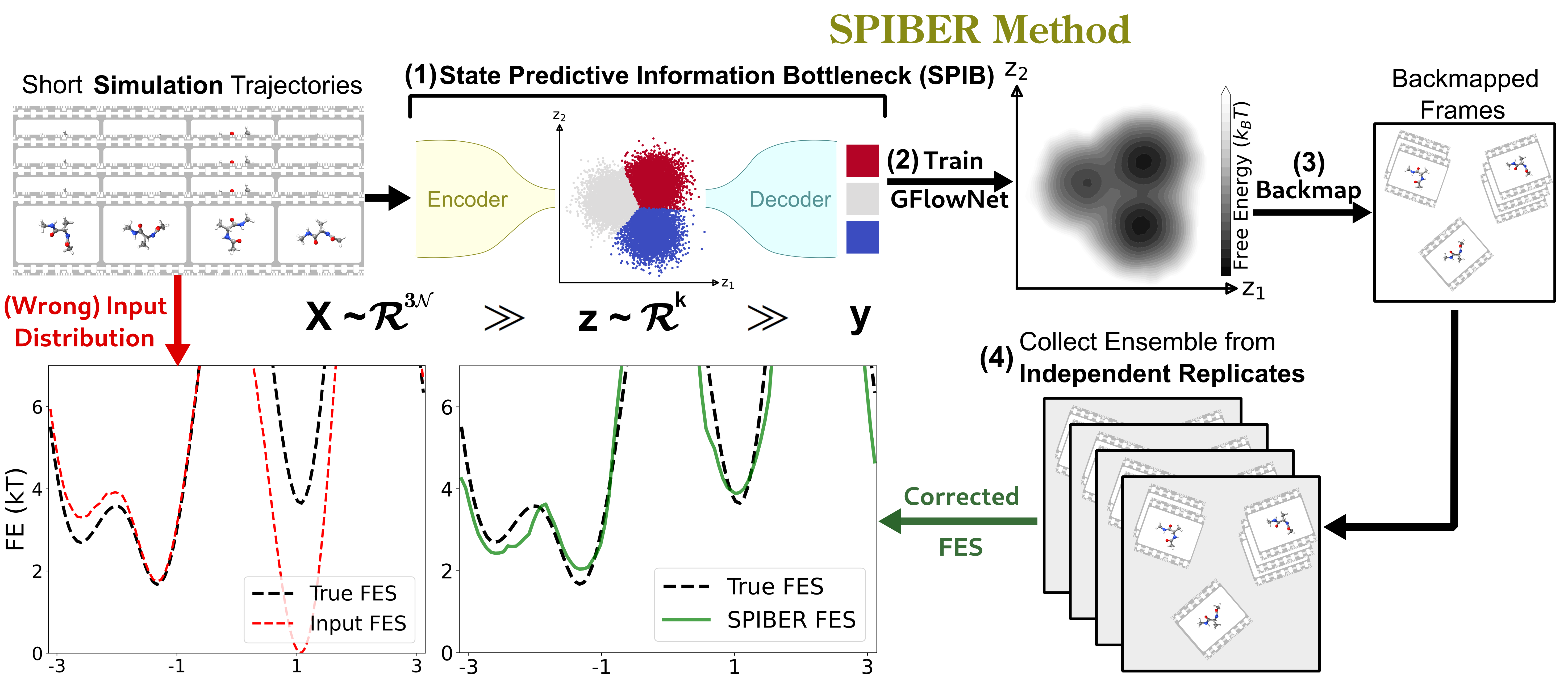}
		\caption{Summarizing different steps of the \methodname~protocol. Given available input data consisting of short, unbiased MD simulations, we (1) Use SPIB as an information compression algorithm to construct an entropy-free, low-dimensional CV $\mfocus{z} : \mathbb{R}^{3N}\to\mathbb{R}^k$. (2) Use $\hat{E}(z)$ as defined in Eq.~\ref{eqn:energy_estimation}, to sample points in $\mfocus{z}$ with the probability distribution $p(z) \sim e^{-\beta \hat{E}(z)}$. (3) Backmap samples in $\mfocus{z}$ to the nearest MD frames. (4) Repeat the SPIB and GFlowNet training processes multiple times to eliminate training error, and aggregate frames from all repeats to calculate the free energy along any feature (here we show $\phi$). Free energies by \methodname~(green plot) are accurate even when the input data distribution (red plot) is far from equilibrium (black)}
		\label{fig:SPIBER_summary}
	\end{figure*}

	We are now in a position to combine the various individual threads introduced so far and discuss the protocol of this work, which we call \methodnamefull, summarized in Fig.~\ref{fig:SPIBER_summary}. In Sec.~\ref{subsec:ideal_CV}, we have shown a theoretical basis to expect well-regularized SPIB latent spaces to behave as entropy-free CVs and in Sec.~\ref{subsec:energy_estimation} and~\ref{subsec:gfnet_sampling} we described how such CVs can be used to perform energy-based reweighting by training GFlowNets only on existing MD data. 
	However, the theoretical framework above and SPIB itself (See Appendix~\ref{appendix:SPIB}) leave several practical decisions to be specified. We describe our choices below.
	
	\subsubsection{Learning an \textit{entropy-free} CV with SPIB}
	As can be seen in Fig.~\ref{fig:SPIBER_summary}, our starting point is a collection of relatively short, typically unconverged simulation trajectories collected in full resolution $\mfocus{x} \in \mathbb{R}^{3N}$. We collate these trajectories and learn a low-dimensional (1-4D) SPIB as a function of input features. While recent work using Graph Neural Networks\cite{zou2025graph} allows learning SPIB directly from all-atom coordinates, here we work with a subset of the input features such as dihedrals. The only requirement of providing this curated input is that all degrees of freedom that can disproportionately affect the entropy between different metastable states must be included. For example, bond vibrations in any molecule contribute almost exactly identically to all states and thus bond lengths can safely be excluded. With this input, SPIB is trained as usual except with high regularization (high $\beta_{IB}$ in Eq.~\ref{eqn:SPIB_loss}). See Appendix~\ref{appendix:SPIB} for more details on this process. Once the SPIB is learnt, our next step is to utilize this entropy-free CV for energy-based reweighting using the potential energy along $\mfocus{z}$ given by Eq.~\ref{eqn:energy_estimation_practical}. Using this definition of $\hat{E}(z)$, sampling points on the latent space, $\mfocus{z}$ with probability $p(z)\sim e^{-\beta\hat{E}(z)}$ reproduces the Boltzmann distribution under the assumptions underlined in Sec.~\ref{subsec:ideal_CV}. This is done here using GFlowNets.
	
	\subsubsection{Training GFlowNets on SPIB CVs using simulation data}
	Traditionally, GFlowNets are designed for discrete object construction. To sample continuous latent coordinates, we follow the established binary discretization protocol where each dimension is represented by 16 bits, and constructing a $k$-dimensional point requires $16k$ sequential binary decisions\cite{EBGFN_paper_PMLR}. After training, sampling is linear in $k$, demonstrating how amortization enables tractable high-dimensional exploration.
	
	Training requires careful balancing. Much of the latent space contains no nearby data, yielding sparse rewards that can cause mode collapse. To prevent this, we construct a balanced training set by sampling approximately 150 points uniformly from each SPIB-predicted state.
	
	GFlowNets can be trained on both forward trajectories (generated by the policy and then scored) and backward trajectories (sampled from a target configuration backward to the root)\cite{TrajectoryBalance}. Forward trajectories represent exploration, while backward trajectories represent exploitation in the Reinforcement Learning framework. Our training proceeds using this TRajectory Balance loss in two phases:
	
	\textit{Equilibration phase.} For the first 30,000 batches (batch size 64), only 20\% of trajectories are backward samples drawn from known data points. The remaining 80\% are forward explorations, allowing the policy to learn the global structure of the latent space with minimal bias.
	
	\textit{Focusing phase.} For the subsequent 30,000 batches, we increase the fraction of backward trajectories to 60\%, anchoring the policy around regions with known populations. This prevents drift while preserving enough forward exploration to cover the full distribution.
	
	Together, these phases transform biased input data into correctly reweighted thermodynamic distributions.

	\subsubsection{Back-mapping to input features/physical observables}
	The GFlowNet generates samples in the SPIB latent space. To connect these back to physically interpretable quantities, which could be the input features used in SPIB or some other features desired by the user of \methodname, we locate the nearest MD frames (in latent-space distance) for each generated sample and use their full atomic coordinates to compute any observable of interest. Specifically, we project our results onto: radial distance $r$ for the double-well particle; backbone dihedral angles $(\phi, \psi)$ for alanine dipeptide; and $\zeta$ (the sum of five internal $\phi$ dihedrals) for \Aib.
	
	\subsubsection{Repeating \methodname~over multiple replicates}
	Because SPIB operates on finite trajectory data, individual training runs can yield slightly different latent spaces depending on the random initialization. To ensure robustness against such variability, we train 5--12 independent SPIB models from distinct random seeds and carry each through the full \methodname~pipeline --- $\hat{E}(z)$ estimation, GFlowNet training, and sampling --- independently. From the resulting ensemble of GFlowNet outputs, we draw an equal number of samples per replicate and pool them into a single backmapped population. The FES is then computed from this combined distribution, which averages out run-to-run fluctuations while preserving the underlying thermodynamic signal.

	\section{Results}
	\label{sec:results}
	We benchmark \methodname~on three systems of increasing complexity: an analytical asymmetric radial double-well potential, alanine dipeptide in vacuum, and the oligopeptide {\Aib} in vacuum. For each system, we demonstrate that SPIB-learned latent spaces enable accurate free energy reweighting via GFlowNets, while conventional CVs (PCA, tICA, and human-chosen dihedrals) fail to produce meaningful corrections. Throughout, we compare against MBAR as a baseline to verify that other similar energy-based frame reweighting approaches cannot achieve the same result. All FESs are optimally aligned to the reference by adding a constant offset.
	
	\subsection{Asymmetric Radial Double-well Potential}
\begin{figure}
		\captionsetup{skip=6pt}
		\centering
		\includegraphics[width=0.4\textwidth]{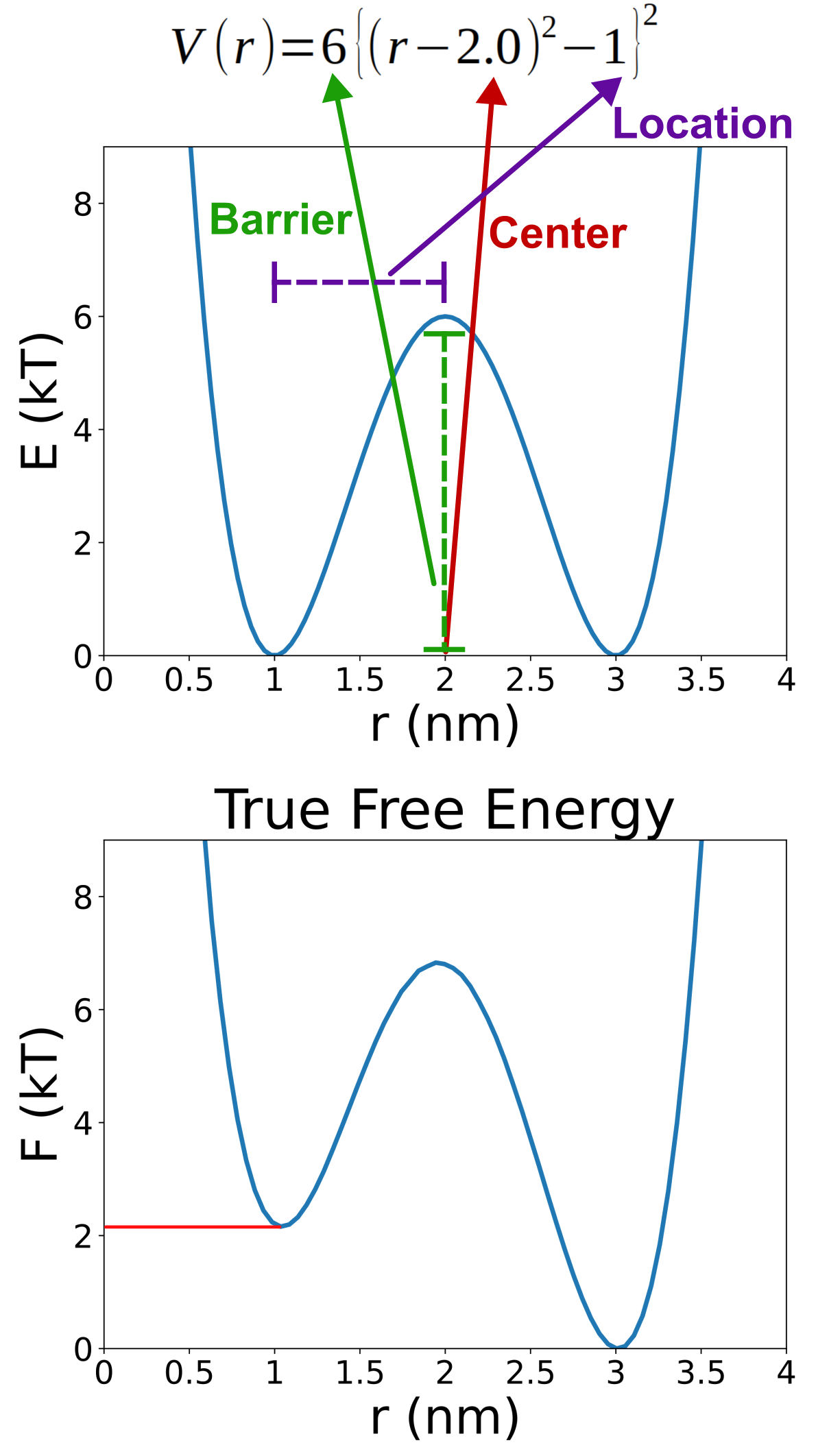}
		\caption{Radial double-well potential and free energy profiles along $r$. The upper panel shows the potential energy $V(r)$ along $r$ --- a symmetric double-well with wells at $r=1$ and $r=3$, and a barrier at $r=2$. The lower panel shows the true free energy computed from an ensemble average of 120 long, independent trajectories, with the horizontal red line roughly indicating the free energy minima value at $r=1$. The outer well ($r=3$) is entropically favored by $\approx 2.19\,k_BT$.}
		\label{fig:Radial_Setup}
	\end{figure}

	As a proof of principle, we simulate a single particle in a radial double-well potential (Fig.~\ref{fig:Radial_Setup}):
	\[V(r) = 6\{(r-2)^2-1\}, \quad r^2 = x^2 + y^2 + z^2.\]
	The two wells sit at $r=1$ and $r=3$. Because the outer well has $\sim 9\times$ more configurational volume, it is entropically favored despite identical potential depths. The expected free energy difference is
	\[F(r=3) - F(r=1) \approx k_BT\ln\left(\frac{3^2}{1^2}\right) \approx 2.19\,k_BT.\]
	
\begin{figure*}
		\captionsetup{skip=6pt}
		\centering
		\includegraphics[width=0.99\textwidth]{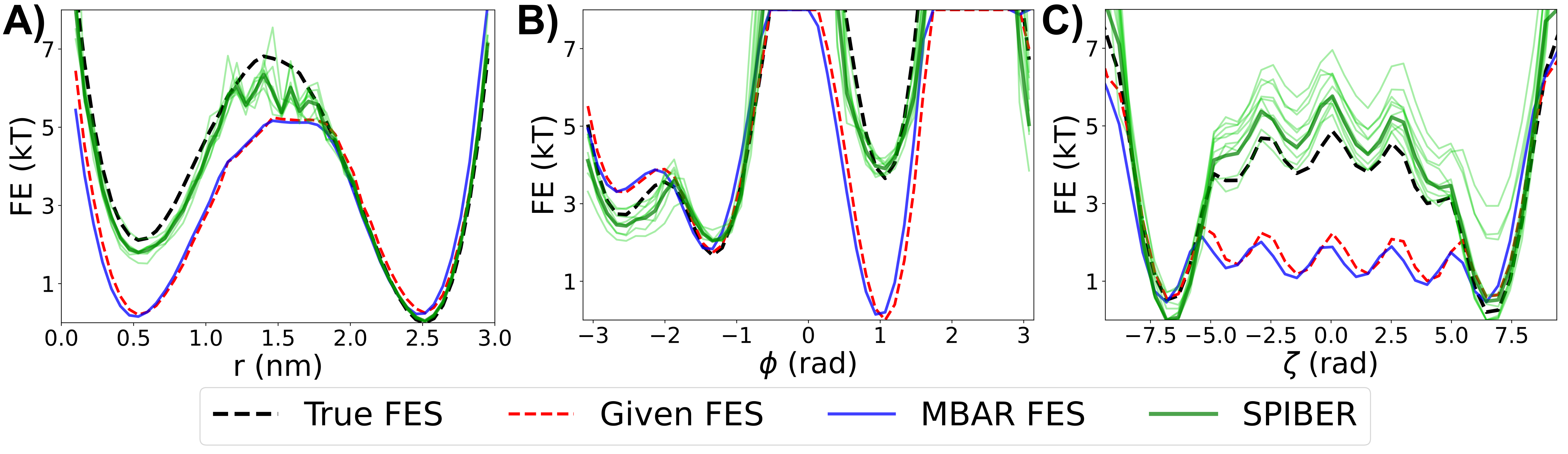}
		\caption{Reweighting results from \methodname~for all systems using SPIB-learned CVs projected along human-interpretable CVs: A) Radial Double-Well along $r$,~B) Alanine Dipeptide along $\phi$~and~C) \Aib~along $\zeta$. The true free energies (black, dashed) near the minima are reproduced by \methodname~(green), despite biased input (dashed, red). MBAR (blue) shows no correction.}
		\label{fig:SPIB_FES_Results}
	\end{figure*}

	We ran 120 independent Langevin dynamics simulations\cite{LangevinDynamics} using the BAOAB integrator\cite{LangevinBAOAB} of 250 time units each ($\gamma=50$, $dt=10^{-4}$, frames every 100 steps). Individual trajectories are short enough that most particles undergo $\leq 2$ transitions between wells, but the ensemble converges to the correct FES and serves as our reference. To test \methodname's reweighting capability, we selected four trajectories with deliberately biased sampling: each had 1--2 transitions, and the combined distribution was nearly uniform across both wells (shown in red in Fig.~\ref{fig:SPIB_FES_Results}A).
	
	SPIB was trained on a 6D input $(x,y,z,x^2,y^2,z^2)$ with a 1024-neuron encoder and decoder, lag time 7.5, and $\beta_{IB}=0.1$ (the highest value before state prediction accuracy dropped). Despite the biased input, \methodname~recovers the correct FES by leveraging equilibrium sampling \textit{within} each metastable state combined with coordinate information (Fig.~\ref{fig:SPIB_FES_Results}A, green). MBAR, applied to the same four trajectories, produces no correction, as expected when all samples come from the same unbiased Hamiltonian at a single temperature.
	
	\subsection{Alanine Dipeptide in Vacuum}
	Alanine dipeptide is a benchmark biomolecular system with three well-characterized metastable states in vacuum at 450 K: $\mathrm{C7_{eq}}$ ($\beta$-sheet-like extended conformation, $\phi \approx -2.5$ rad), $\mathrm{C_5}$ (extended 5-member ring state, $\phi \approx -1.9$ rad), and $\mathrm{C7_{ax}}$ (left-handed helical turn, $\phi \approx +1$ rad)\cite{ALA2_States}. The ground-truth free energy surface from a 1~$\mu$s MD simulation is shown in Fig.~\ref{SI-FES_Ala2} of the Supporting Information. To test reweighting from biased input, we selected 12 contiguous segments from the long trajectory that over-represent $\mathrm{C7_{ax}}$ and under-represent the true global minima.
	
	\subsubsection{Using SPIB in 2D}
	As a first test, SPIB was trained with an 8D input (sine and cosine of $\phi$, $\psi$, $\theta$, $\omega$), a 256-neuron encoder and decoder, lag time 0.3 ps, and $\beta_{IB}=0.25$. Initial state labels followed our original SPIB protocol\cite{SPIBOriginal} (10 states based on $\phi$). The short lag time captures rapid transitions between $\mathrm{C7_{eq}}$ and $\mathrm{C_5}$. As shown in Fig.~\ref{fig:SPIB_FES_Results}B, \methodname~faithfully reproduces the free energies of all three minima along $\phi$, despite the artificially biased input. MBAR again shows no correction capacity. A similar comparison projected onto $\psi$ is shown in Fig.~\ref{SI-PSI-ALA2} of the Supporting Information.
	
	\subsubsection{Using known collective variables}
	As a control, we tested whether human-chosen CVs work for energy-based sampling with GFlowNets. The standard choice of backbone dihedrals $\phi$ and $\psi$ accounts for nearly all relevant conformational transitions in Alanine Dipeptide. With these as latent variables, most relevant degrees of freedom are captured, and reweighting near metastable states is reasonable (Fig.~\ref{SI-FES_Ala2_PhiPsi}A of the Supporting Information), but this setup in energy-based sampling still fails to perform as well as \methodname.
	
	Replacing $\phi$ with $\theta$ as the second CV ($\psi,\theta$) leads to complete failure: since entropy now depends non-trivially on the unobserved $\phi$, reweighting cannot recover correct free energy differences even at local minima (Fig.~\ref{SI-FES_Ala2_PhiPsi}B of the Supporting Information). This illustrates that not all state-distinguishing CVs are suitable for energy-based reweighting.
	
	\subsection{Conformers of \Aib}
\begin{figure}[h]
		\captionsetup{skip=6pt}
		\centering
		\includegraphics[width=0.45\textwidth]{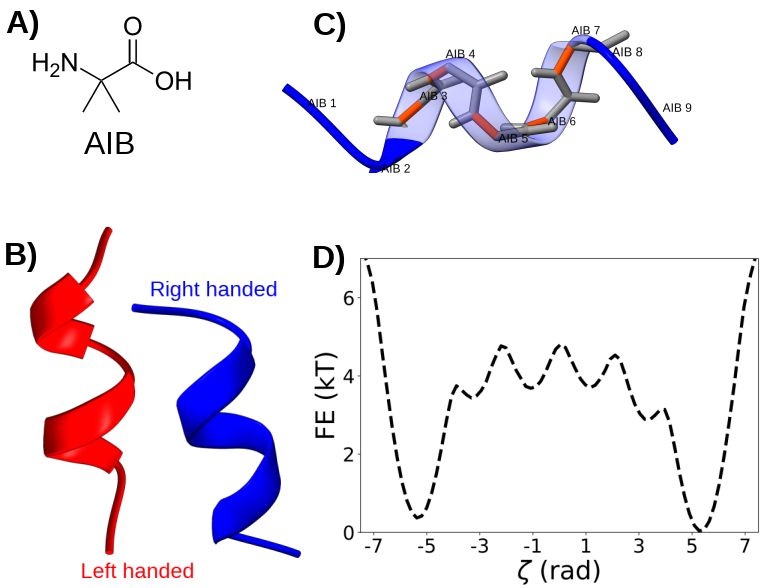}
		\caption{\Aib~overview: A) Chemical structure of AIB, B) Right- and left-handed helical forms, C) Definition of $\zeta$ (sum of five internal dihedrals, bonds highlighted in orange), D) Ground-truth FES along $\zeta$ from 2.5~$\mu$s MD at 500 K.}
		\label{fig:about_AIB9}
	\end{figure}

	\Aib~is a 9-residue oligopeptide composed entirely of non-chiral 2-aminoisobutyric acid (Fig.~\ref{fig:about_AIB9}A). It forms two stable helical structures, right-handed and left-handed, with multiple metastable intermediates containing partial helices. The standard CV is $\zeta$, defined as the sum of five internal $\phi$ dihedrals (Fig.~\ref{fig:about_AIB9}C), which peaks at $+5$ and $-5$ for complete right- and left-handed helices, respectively\cite{biswas2018metadynamics,AIB9_energy_transport}. Transitions between these forms are rare at 300 K, so we obtained the ground-truth FES from a 2.5~$\mu$s simulation at 500 K (Fig.~\ref{fig:about_AIB9}D).
	
	\subsubsection{Using $\zeta$ as a 1D CV}
	
	Unlike $\phi,\psi$ for Alanine Dipeptide, $\mfocus{\zeta}$ is a degenerate CV: many distinct conformations share the same $\mfocus{\zeta}$ value. While it separates metastable states qualitatively, the large conditional entropy $H(\mfocus{X}|\mfocus{\zeta})$ violates the assumption behind Eq.~\ref{eqn:spiber_approx}. As shown in Fig.~\ref{SI-AIB9_Results_Zeta} of the Supporting Information, reweighting with $\zeta$ alone fails to recover the correct FES. This reinforces a broader point: most human-designed CVs excel at state discrimination but are unsuitable for free energy estimation.
	
	\subsubsection{Using SPIB in 3D}
	
	Following previous work on \Aib~\cite{ShamsSPIB}, we binarize the five internal dihedrals (left- vs.\ right-handed, $\phi > 0$ or $\phi < 0$), yielding 32 initial states. SPIB isolates 15--16 stable states depending on $\beta_{IB}$ (the state count as a function of $\beta_{IB}$ is shown in Fig.~\ref{SI-BETA-TUNING}C of the Supporting Information). We used a lag time of 100 ps, a 512-neuron encoder and 256-neuron decoder.
	
	Due to sparse sampling even at 500 K, we lowered the neighborhood threshold, accepting regions with just 42 points. As shown in Fig.~\ref{fig:SPIB_FES_Results}C, \methodname~recovers free energies close to the ground truth across all resolved states.
	
	\subsubsection{Using SPIB in 4D}
\begin{figure}
		\captionsetup{skip=6pt}
		\centering
		\includegraphics[width=0.45\textwidth]{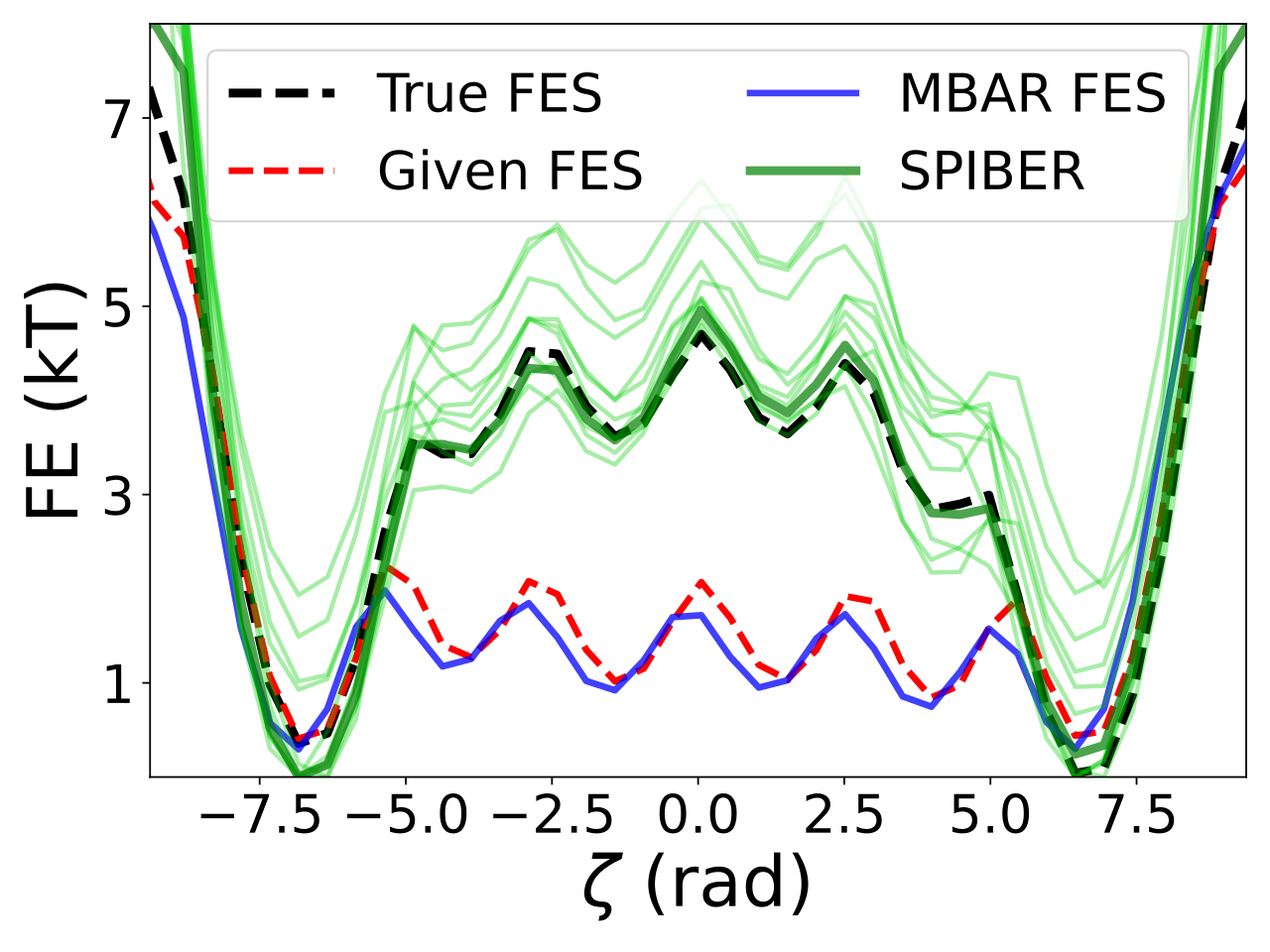}
		\caption{Reweighting results for \Aib~using SPIB-learned 4D CVs. The much improved accuracy over the 3D case highlights the benefit of higher-dimensional latent spaces when GFlowNets are used for sampling.}
		\label{fig:AIB9_Results_SPIB_4D}
	\end{figure}

	To test whether higher-dimensional latent spaces improve results, we repeated the \Aib~analysis with a 4D SPIB latent space (all other parameters unchanged). The more expressive representation does yield improved accuracy (Fig.~\ref{fig:AIB9_Results_SPIB_4D}), demonstrating that GFlowNets remain tractable in higher dimensions where traditional Monte Carlo methods struggle. However, beyond 4D the effective volume of latent space grows rapidly, making $\hat{E}(z)$ estimates unreliable due to sparse MD frame coverage. Reweighting with a 5D latent variable obscures sparsely populated intermediates: projecting a fixed number of frames into a higher-dimensional space increases the total volume without adding data, making the latent representation sparser. Performance may improve with more samples, though we have not validated this here.
	
	\subsection{Comparison with PCA and tICA}
\begin{figure}
		\centering
		\includegraphics[width=0.4\textwidth]{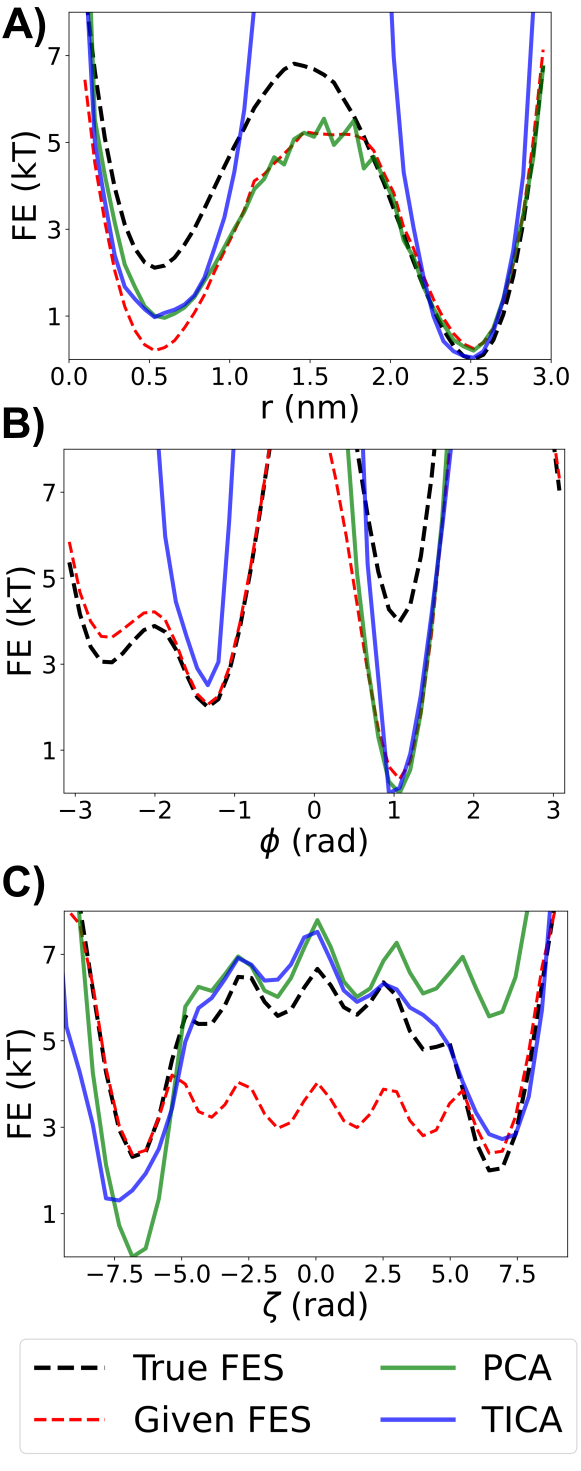}
		\caption{Results using PCA and tICA for A) Radial Double-Well B) Alanine Dipeptide and C) \Aib. PCA begins missing states as the dimensionality increases. tICA performs better, but it is still inconsistent.}
		\label{fig:FES_PCA_TICA}
	\end{figure}

	To confirm that the quality of the latent space --- not the GFlowNet sampler --- determines reweighting success, we repeated all three benchmarks using PCA and tICA as CVs instead of SPIB. Inputs, lag times (where applicable), and GFlowNet hyperparameters were held identical across methods. In every case, PCA and tICA failed to recover meaningful free energy corrections: entropically favored states had average energies too high to be sampled by the GFlowNet, causing entire minima to disappear from the reweighted distribution. As shown in Fig.~\ref{fig:FES_PCA_TICA}, neither PCA nor tICA could accurately reproduce the free energies for the metastable states. PCA performs the worst, failing almost immediately while tICA is relatively better, managing to approximately recover the intermediate states for \Aib, but missing the global minimum by more than $1~k_BT$, and completely missing a metastable state for alanine dipeptide. Neither method outperforms \methodname~on any system.
	These results are consistent with the theoretical analysis in Sec.~\ref{subsec:CVrules}: PCA maximizes $I(\mfocus{X};\mfocus{z})$ (the opposite of what SPIB does), and fails almost instantly once dimensionality increases even slightly. tICA performs like unregularized SPIB, producing sharp state boundaries which cannot be used for accurate reweighting.
	
	\subsection{Comparison with Other Reweighting Methods}
	\label{subsec:results-other-methods}
	MBAR was applied alongside \methodname~in all three benchmarks. In every case, MBAR produced no correction beyond the raw histogram (blue curves in Fig.~\ref{fig:SPIB_FES_Results}). This is expected: MBAR requires either Hamiltonian perturbations or substantial overlap between state populations, neither of which holds for short unbiased trajectories at a single temperature.
	
	We also tested DHAM (Dynamic Histogram Analysis Method)\cite{DHAM}, which estimates free energies from transition counts between states along a collective variable. Because our trajectories contain no added bias, DHAM effectively reduces to fitting a Markov state model and extracting free energies from the detailed balance condition. The MSM was constructed by dividing the observable CV ($r$ for the radial double-wells, $\phi$ for Alanine Dipeptide, and $\zeta$ for \Aib) into 100 equal-width bins; the initial transition matrix was built from raw transition counts and then iteratively converged following the DHAM procedure\cite{DHAM_Iterative}. As shown in Fig.~\ref{SI-DHAM} of the Supporting Information, DHAM yields only marginal improvement over the raw histogram, and the results are inconsistent across systems --- for \Aib, the estimate is actually degraded due to the oversampling of transitions inherent in our short-trajectory design. We found no improvement even on changing the number of states for DHAM. Even under more realistic sampling conditions (radial double-wells and Alanine Dipeptide), DHAM fails to meaningfully correct population imbalances. To our knowledge, no existing method consistently corrects population imbalances in such data.
	
	\section{Conclusions and Discussion}
	\label{sec:Conclusion}
	Short molecular dynamics trajectories can reveal several metastable conformations while providing a misleading picture of their relative thermodynamic stability. Here, we developed \methodname, an energy-based sampling method for reconstructing free energy landscapes from short, unconverged trajectories. \methodname~combines collective variables learned through the State Predictive Information Bottleneck (SPIB), conditional mean potential energies, and Generative Flow Networks (GFlowNets), without requiring additional molecular dynamics. The key idea is that energy-based sampling does not require converged global state populations when conditional mean potential energies can be estimated accurately along an entropy-free CV. For such a CV, conditional entropy is approximately independent of position, allowing these energies to determine approximate equilibrium probabilities. We show how well-regularized SPIB CVs can naturally achieve this property by compressing configurational information while preserving predictions of future metastable states.
	
	Across a radial double-well model, alanine dipeptide, and \((\mathrm{Aib})_9\), \methodname~improves projected free energy profiles, particularly near represented minima. It is also able to scale to four-dimensional latent spaces, a regime in which traditional Monte Carlo sampling becomes computationally prohibitive\cite{MonteCarloReview}. The 4D representation of \(\mathrm{(Aib)}_9\) shows marked improvement over its 3D counterpart, and the framework likely remains feasible in even higher dimensions given sufficient sampling coverage. Comparisons with PCA, tICA, and manually selected CVs demonstrate the importance of the SPIB-based representation for this energy-based estimator.
	
	A promising future application is to ensembles obtained from short MD simulations initialized from diverse AI-predicted structures\cite{Jumper_AlphaFold,RNAnneal,Boltz2,Chai1}. These simulations can explore different regions of the conformational landscape without establishing their relative equilibrium populations, providing a natural setting for \methodname. Further validation, particularly for larger biomolecules and explicit solvent, will be needed to establish its accuracy in this setting.
	
	\section*{Supporting Information} 
	The Supporting Information contains details on the \methodname algorithm, such as the choice of neighborhood radius, minimum sample threshold, and $\beta_{IB}$, as well as hyperparameter values for all benchmark systems, and stability tests for the FES predictions. It also reports other results with projections of the Ala2 FES along $\psi$, comparisons with DHAM reweighting, and demonstrations of energy-based reweighting using hand-chosen collective variables.
	
	\section*{Acknowledgments}
	This research was entirely supported by the US Department of Energy, Office of Science, Basic Energy Sciences, CPIMS Program, under Award DE-SC0021009. We thank UMD HPC’s Zaratan and NSF ACCESS (project CHE180027P) for computational resources. P.T. is an investigator at the University of Maryland-Institute for Health Computing, which is supported by funding from Montgomery County, Maryland, and the University of Maryland Strategic Partnership: MPowering the State, a formal collaboration between the University of Maryland, College Park, and the University of Maryland, Baltimore.\\
	We also thank Nicholas Hattrup and Dedi Wang for their insightful comments and careful review of the manuscript.

	\section*{Code availability statement}
	All the code used for this work along with an easy-to-implement solution for a general system will be made available on our group GitHub before formal submission to a journal for peer review.
	
	\section*{Conflict of Interest}
	The authors have no conflict of interest to report. \newline
	
	\section*{References}
	\bibliography{references}
	
	\makeatletter
	\begingroup
	\renewcommand{\thesubsection}{A.\arabic{subsection}}
	\def\p@subsection{} 
	
	\section*{Appendix}
	\setcounter{subsection}{0} 
	\subsection{Proof that the constraint $\mfocus{H(X|z=z_i) \leq H(X)}$ holds as $\mfocus{I(\mfocus{X};\mfocus{z}) \to 0}$}
	\label{appendix:proof}
	Since $\mfocus{z}$  is a deterministic function of $\mfocus{X}$, the conditional entropy $H(\mfocus{z}|\mfocus{X}) = 0$, which gives
	\begin{equation}
		\label{eqn:mutual_information_twoway}
		\epsilon = I(\mfocus{X};\mfocus{z}) = H(\mfocus{z}) - H(\mfocus{z}|\mfocus{X}) = H(\mfocus{z}) \geq 0
	\end{equation}
	since mutual information is non-negative.\\
	Define deviations from the mean conditional entropy
	\begin{equation}
		\label{eqn:deviation_defn}
		d_i\equiv\mu-h_i,
		\qquad
		h_i\equiv H(\mfocus{X}\mid\mfocus{z}=z_i),
		\qquad
		\sum_{i=1}^{M}p(z_i)d_i=0.
	\end{equation}
	Since $h_i\geq0$, we have $d_i\leq\mu$, and for any region with
	$p(z_i)>0$,
	\begin{equation}
		\label{eqn:deviation_sum}
		0=p(z_i)d_i+\sum_{j\neq i}p(z_j)d_j
		\leq p(z_i)d_i+\mu[1-p(z_i)],
	\end{equation}
	which gives
	\begin{equation}
		\label{eqn:deviation_limit}
		\forall i,~~d_i\geq\mu\left[1-\frac{1}{p(z_i)}\right] \Longrightarrow \ln\left(\frac{h_i}{\mu}\right)\leq-\ln p(z_i).
	\end{equation}
	using $d_i \equiv \mu -h_i$ as defined in Eq.~\ref{eqn:deviation_defn}\\
	Let
	\[
	\mathcal{K}
	\equiv
	\left\{i:
	h_i\geq H(\mfocus{X})\right\}
	\]
	denote the regions that violate the upper bound. Since $H(\mfocus{X})=\mu+\epsilon$, each term
	$\ln(h_i/\mu)$ is nonnegative for $i\in\mathcal{K}$. A $p(z)$-weighted sum of Eq.~\ref{eqn:deviation_limit} over these regions gives
	\begin{equation}
		\label{eqn:log_deviation_limit}
		0\leq
		\sum_{i\in\mathcal{K}}p(z_i)\ln\left(\frac{h_i}{\mu}\right)
		\leq
		-\sum_{i\in\mathcal{K}}p(z_i)\ln p(z_i)
		\leq H(\mfocus{z})=\epsilon.
	\end{equation}
	Consequently, for every $i\in\mathcal{K}$,
	\begin{equation}
		\label{eqn:violating_deviation_limit}
		h_i
		\leq
		\mu\exp\left[\frac{\epsilon}{p(z_i)}\right].
	\end{equation}
	For any region satisfying $\epsilon/p(z_i)\to0$, this gives
	\[
	0\leq h_i-H(\mfocus{X})
	\leq
	\mu\exp\left[\frac{\epsilon}{p(z_i)}\right]
	-(\mu+\epsilon)
	\longrightarrow0.
	\]
	This condition holds whenever $p(z_i)$ remains nonzero as $\epsilon\to0$, and more generally whenever $p(z_i)$ vanishes more slowly than $\epsilon$. Thus, upper-bound violations vanish throughout the populated regions of CV space.\\
	No pointwise conclusion follows for rare regions with $p(z_i)=O(\epsilon)$ or smaller; however, their combined contribution to the sampling domain vanishes faster than $O(\epsilon^2)$, namely, 
	\[
	\sum_{i\in\mathcal R}p(z_i)\sim O(\epsilon^2),
	\qquad\text{whenever}\qquad
	\sum_{i\in\mathcal R}p(z_i)(h_i-\mu)^2 >0
	\]
	for any collection $\mathcal R$ of rare regions.
	
	\subsection{State Predictive Information Bottleneck method}
	\label{appendix:SPIB}
	
	State Predictive Information Bottleneck (SPIB)~\cite{SPIBOriginal} is a deep learning framework for extracting low-dimensional collective variables from high-dimensional trajectory data and is particularly suited for \methodname because part of its optimization objective --- minimizing $I(\mfocus{X};\mfocus{z})$ subject to predictive constraints --- directly produces the near-constant conditional entropy required for energy-based reweighting (Sec.~\ref{subsec:CVrules}).
	
	\paragraph{Optimization Objective.}
	Based on Rate-distortion theory\cite{Shannon1948, Alemi_ICLR}, SPIB formulates CV learning as a constrained information bottleneck problem\cite{DVIB}. Its loss function is given by Eq.~\ref{eqn:SPIB_loss} in the main text. The first term $I(\mfocus{y};\mfocus{z})$ ensures that the latent representation retains predictive power for future metastable states, making it a physically meaningful CV. The second term $I(\mfocus{X};\mfocus{z})$, weighted by $\beta_{IB}$, compresses away information about instantaneous configurations that is irrelevant to state prediction.
	
	\paragraph{Variational Implementation.}
	Following the Deep Variational Information Bottleneck~\cite{DVIB}, SPIB parameterizes the encoder as a Gaussian:
	\begin{equation}
		\label{eqn:SPIB_encoder}
		p_\theta(\mfocus{z}|\mfocus{X}) = \mathcal{N}\bigl(\bm{\mu}_\theta(\mfocus{X}),\, \text{diag}(\bm{\sigma}_\theta(\mfocus{X}))\bigr)\;,
	\end{equation}
	where $\bm{\mu}_\theta$ and $\bm{\sigma}_\theta$ are outputs of fully connected neural networks. The reparameterization trick~\cite{KingmaWelling2013} enables gradient-based optimization: $\mfocus{z} = \bm{\mu}_\theta(\mfocus{X}) + \bm{\sigma}_\theta(\mfocus{X}) \odot \bm{\epsilon}$ with $\bm{\epsilon} \sim \mathcal{N}(\bm{0},\bm{I})$. The decoder is a softmax classifier that predicts state-label probabilities from the latent code:
	\begin{equation}
		\label{eqn:SPIB_decoder}
		p_\phi(\mfocus{y}|\mfocus{z}) = \text{softmax}\bigl(\mathbf{W}_\phi\, \mfocus{z} + \bm{b}_\phi\bigr)\;,
	\end{equation}
	where $\mfocus{y}$ is one-hot encoded over $D$ discrete states, and the mutual information terms in Eq.~\ref{eqn:SPIB_loss} are estimated variationally. The latent-prior used for SPIB is multi-Gaussian $r_\theta(\mfocus{z})$, constructed via the VampPrior technique~\cite{Tomczak2017Vamp}.
	
	\paragraph{Iterative State Refinement.}
	A distinctive feature of SPIB is its iterative self-training scheme, which eliminates the need for a priori knowledge of state identities. The algorithm proceeds as follows:
	
	\begin{enumerate}
		\item \textbf{Initialization:} Assign arbitrary state labels $\{\mfocus{y}^n\}$ to trajectory frames. This can be done by discretizing an intuitive CV (e.g., binning a dihedral angle), or by assigning random labels.
		\item \textbf{Training cycle:} Optimize the loss using mini-batch gradient descent.
		\item \textbf{State relabeling:} After each training cycle, update state labels using the decoder's output. This relabeling step also merges configurations with similar dynamical futures and discards labels for states that no longer have support.
		\item \textbf{Convergence:} Repeat steps 2--3 until the latent representation, state-transition density, and discrete state labels stabilize. The converged number of states is determined self-consistently by the data and the lag-time $\Delta t$.
	\end{enumerate}
	This iterative design means that SPIB automatically discovers both the number and location of metastable states from trajectory dynamics alone, with minimal human intervention.
	
	\paragraph{Role of the Lag Time $\Delta t$.}
	The lag time $\Delta t$ is the principal user-controlled hyperparameter and serves as a \textit{time resolution} for the coarse-grained description. Physically, it exploits timescale separation: if $t_{\text{fast}} \ll \Delta t \ll t_{\text{slow}}$, where $t_{\text{fast}}$ is the intra-state relaxation time and $t_{\text{slow}}$ is the inter-state transition time, then $\Delta t$ filters out fast fluctuations while preserving slow transitions. A shorter $\Delta t$ resolves finer state structure; a longer $\Delta t$ merges rapidly exchanging substates into broader metastable basins.
	
	\paragraph{Network Architecture and Training Details.}
	For all benchmarks in this work, both encoder and decoder use fully connected networks with two hidden layers and ReLU activation~\cite{ReLU}. Layer widths are system-dependent (see main text for specifics). Training uses the Adam optimizer~\cite{AdamOpt} with learning rate $10^{-5}$ and batch size 256. The hyperparameter $\beta_{IB}$ is tuned by monitoring the number of resolved states as a function of regularization strength; we select the largest $\beta_{IB}$ before state count drops sharply (Fig.~\ref{SI-BETA-TUNING} of the Supporting Information), ensuring maximal compression compatible with faithful state separation.
	
	\endgroup
	\makeatother
\end{document}


\maketitle
	
	\section{Methods}
	The following analyses were used to self-consistently determine any tunable hyperparameters for this method. All adjustable parameters can be determined without any prior knowledge beyond a rough understanding of the state classifications for the system of interest.
	
	\subsection{\methodname~algorithmic details}
	The \methodname~pipeline involves a few tunable choices: the neighborhood radius used to estimate $\hat{E}(z)$, the minimum sample threshold for numerical stability, and the SPIB regularization strength $\beta_{IB}$ along with model size. Below we describe how each is determined self-consistently from the data alone, without requiring prior knowledge of the system.
	\subsubsection{Defining the neighborhood of a point}
\begin{figure*}
		\includegraphics[width=0.8\linewidth]{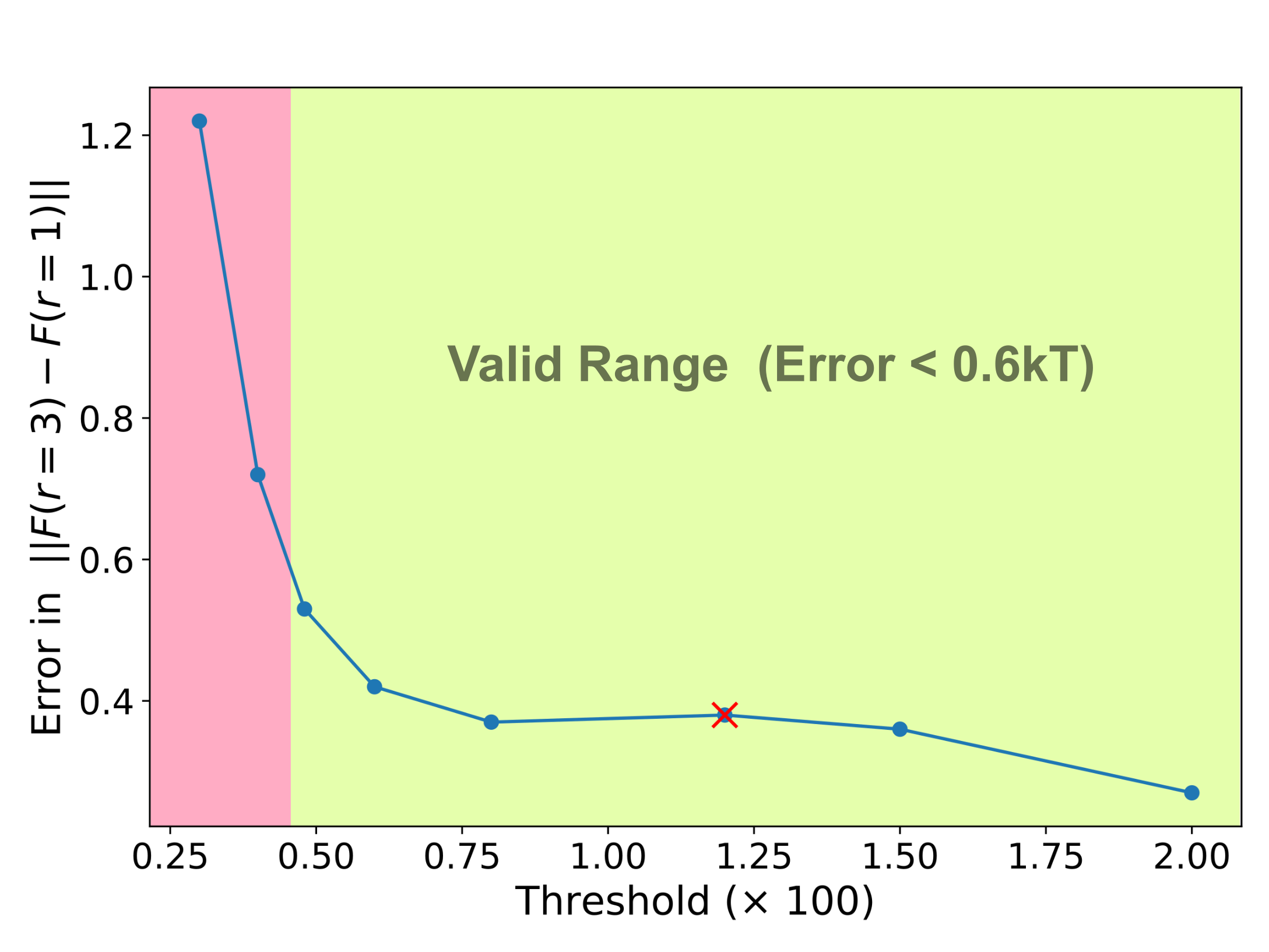}
		\caption{Error in the predicted free energy difference between $r=1$ and $r=3$ wells. The selected threshold ($D/50$, as described in the main manuscript) is marked with a red cross.}
		\label{SI-THRESHOLD-TUNING}
	\end{figure*}

	Since MD frames will almost never share identical latent coordinates, the estimator requires a finite neighborhood around each query point $z^*$ as shown in Eqn~12 of the main manuscript. We define this neighborhood as a hypersphere of radius $R_0$ in the $k$-dimensional latent space. To avoid manual tuning, we set
	\begin{equation}
		\label{eqn:choice_of_r0}
		R_0 = \frac{D}{50}, \quad \text{where} \quad D = \left(\sum_{d=1}^k L_d^2\right)^{1/2},
	\end{equation}
	and $L_d$ is the range (maximum \textit{minus} minimum) of dimension $d$. An example is graphically shown in Figure~\ref{SI-Neighborhood_picking} below. The quantity $D$ corresponds to the length of the longest diagonal of the axis-aligned bounding box spanning the latent space, providing a natural scale for what constitutes a neighborhood. For all systems studied here, $R_0 = D/50$ yields stable results; we additionally demonstrate robustness by varying this denominator in Fig. \ref{SI-THRESHOLD-TUNING} later.
	\begin{figure}
		\captionsetup{skip=6pt}
		\centering
		\includegraphics[width=0.9\textwidth]{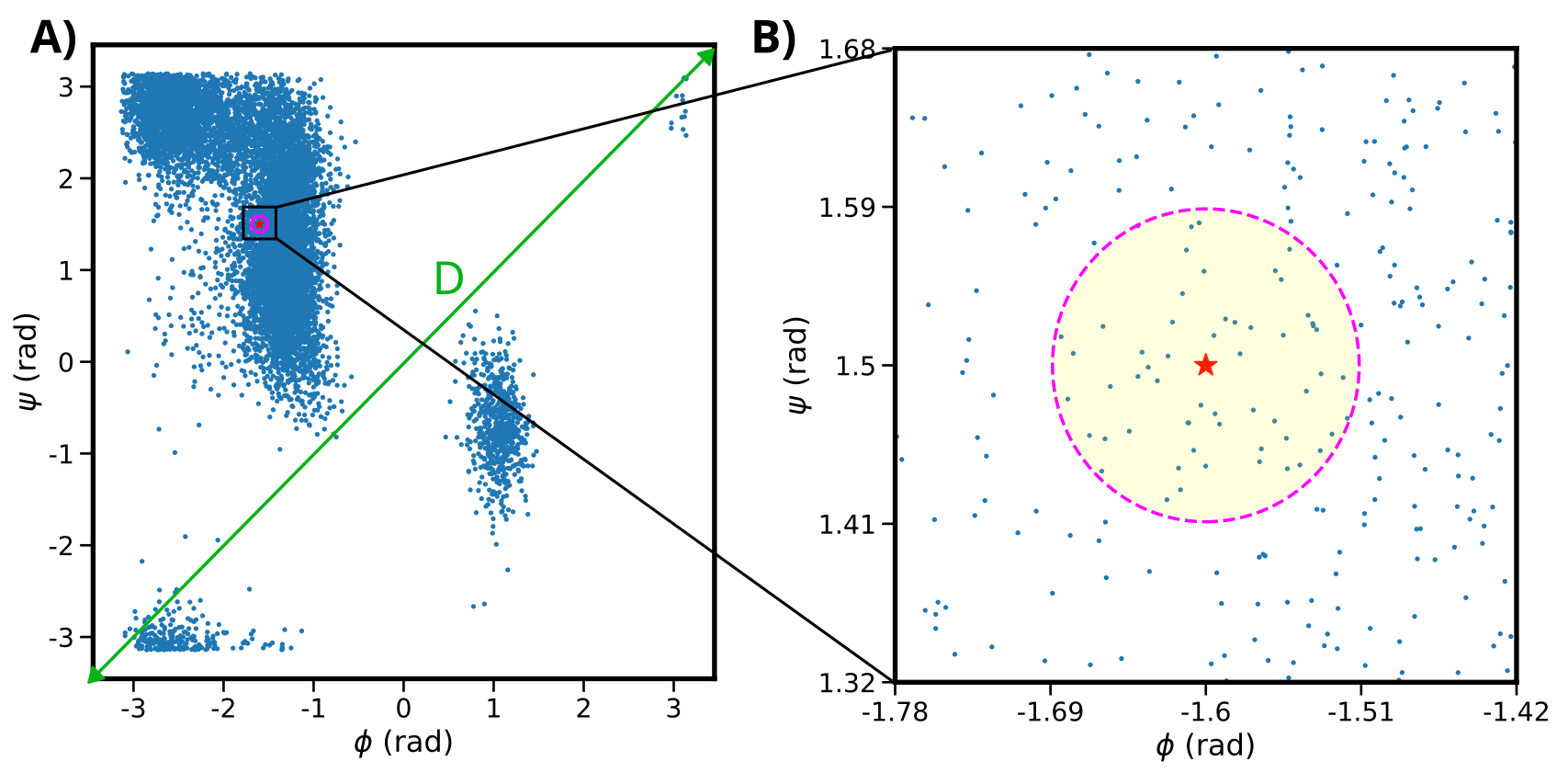}
		\caption{Practical energy-based sampling depicted on Alanine dipeptide's $\phi,\psi$ plot for simplicity. (A) While training, points (red star) are generated randomly in CV space. A small neighborhood (magenta circle) around each such point is considered. $D$ is the longest diagonal as shown by a green line. The radius of the magenta circle is $D/50$ (B) A zoomed-in depiction of a neighborhood of a selected point. All points inside the shaded area are used to calculate the average energy. If there aren't enough points the energy is set to $\infty$}
		\label{SI-Neighborhood_picking}
	\end{figure}
	
	\subsubsection{Minimum sample threshold}
	To maintain numerical stability during GFlowNet training, regions with insufficient data must be excluded from the energy landscape. We set $\hat{E}(z) = \infty$ whenever fewer than 512 simulation frames fall within the $R_0$ neighborhood of $z$. For systems where even this threshold is too stringent (e.g., when a metastable state is poorly represented with fewer than $N_{min}=100$ points), we fall back to using a neighborhood with just 42 or more points, where $N_{\min}$ is the frame count of the least-populated SPIB-predicted state. Higher thresholds improve fidelity but risk excluding valid regions when input data is sparse; our choice represents a practical compromise.
	
	\subsubsection{Choosing $\beta_{IB}$ and model size}
	The entire method hinges on SPIB driving $I(\mfocus{X};\mfocus{z})$ sufficiently close to zero. This requires large values of $\beta_{IB}$, but the optimal value is system-dependent. The trade-off in the SPIB loss, $\mathcal{L}_{IB} \equiv I(\mfocus{z};\mfocus{y}) - \beta_{IB}\, I(\mfocus{X};\mfocus{z})$, implies that excessive regularization will degrade state-prediction accuracy: if too much information is compressed away, $z$ can no longer distinguish metastable states. We exploit this by gradually increasing $\beta_{IB}$ and monitoring the number of distinct states predicted by SPIB. For all test systems (Fig.~\ref{SI-BETA-TUNING} below), there is a sharp drop in the predicted state count once $\beta_{IB}$ crosses a critical threshold. We select the largest $\beta_{IB}$ just before this transition --- the maximum regularization compatible with faithful state separation.
	
	Achieving strong compression sometimes requires larger neural networks. Counterintuitively, the smallest system (the 3D double-well) required the largest model, likely because compressing already low-dimensional data to near-zero mutual information represents a more constrained optimization problem. Table \ref{SI-TABLE-PARAMS} below summarizes the model sizes and $\beta_{IB}$ values used for each system. In all cases, a 2-layer MLP encoder and decoder are used with ReLU as the activation function.
	\newpage
	
	\subsection{Stability of predicted FES over different neighborhood thresholds}
	To compute the average potential energy of the system at a certain value of the CV, we define a neighborhood radius $R_0$ such that 
	\begin{equation}
		\label{eqn:energy_estimation_practical_SI}
		\hat{E}(\mfocus{z}^*) \approx \mathbb{E}_{x \sim \text{data}}\left[ E(\mfocus{x}) \;\big|\; |z(\mfocus{x}) - \mfocus{z}^*|^2 \leq R_0^2 \right].
	\end{equation}
	We test stability against variations in $R_0$ by evaluating multiple thresholds on our first test system, the radial double-well potential:
	\[V(r) = 6\{(r-2)^2-1\}, \quad r^2 = x^2 + y^2 + z^2.\]
	The potential energy is radially symmetric and has minima at $r=1$ and $r=3$, which have a total free energy difference of $ln(9) \approx 2.19kT$ due to volume differences alone.
	\newpage
	
	\subsection{Choice of best $\beta_{IB}$}
	For the free energy approximation ($F(z) \approx \hat{E}(z) + C$) to hold, we require $I(\mfocus{X}, \mfocus{z})$ to be small, achieved by strong regularization of the SPIB loss. However, excessive regularization degrades state prediction, rendering the resulting CV useless. To identify an optimal trade-off, we scan increasing $\beta_{IB}$ values and select the value just before the number of predicted states suddenly drops.
	\begin{figure*}[h]
		\includegraphics[width=1.0\linewidth]{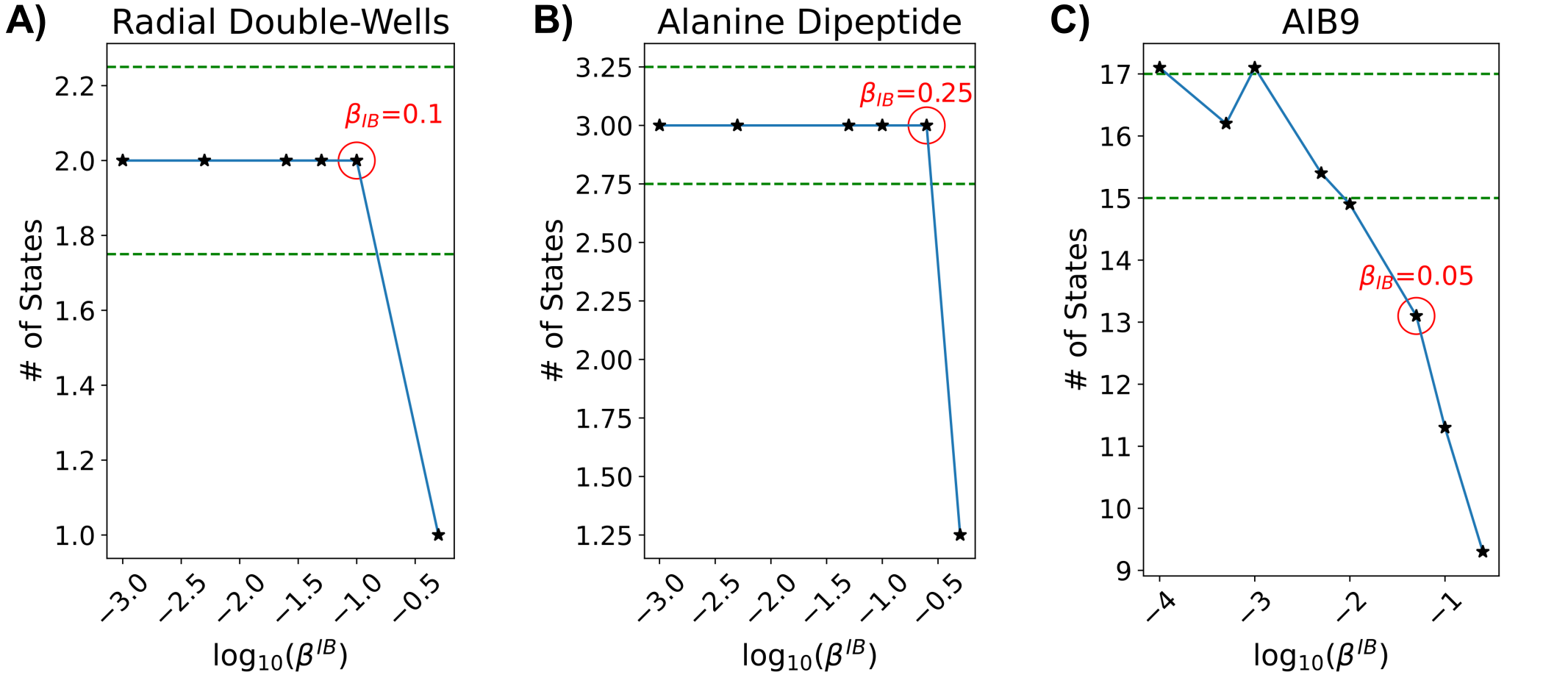}
		\caption{$\beta_{IB}$ was varied from $10^{-4}$ to $0.5$. At low values, state prediction dominates, yielding the best state separation. At high values, the number of states sharply drops, as the least confident states merge.}
		\label{SI-BETA-TUNING}
	\end{figure*}
	In the first two systems shown, excessive regularization collapses all states into a single cluster, which is uninformative. We therefore use the $\beta_{IB}$ value just before this collapse. 
	\newpage
	
	\subsection{Metastable state classification for Alanine Dipeptide}
	Shown below is the ground truth free energy surface of Alanine Dipeptide in $\phi,\psi$ space. The three metastable states identified by SPIB are highlighted and labelled.
	\begin{figure}
		\captionsetup{skip=6pt}
		\centering
		\includegraphics[width=0.67\textwidth]{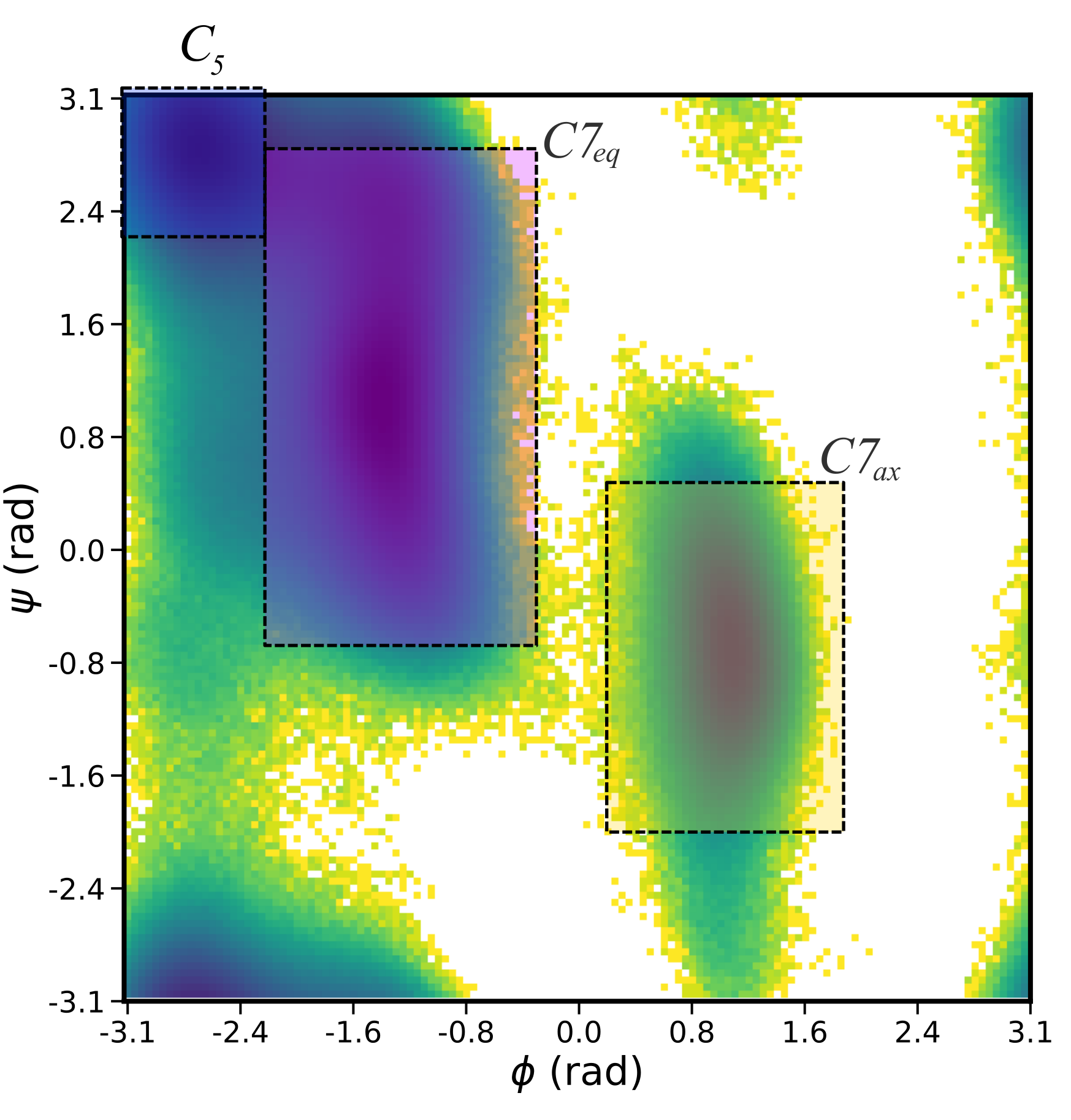}
		\caption{Ground-truth free energy surface of Alanine Dipeptide at 450 K, showing the three metastable minima: $\mathrm{C7_{eq}}$, $\mathrm{C_5}$, and $\mathrm{C7_{ax}}$.}
		\label{SI-FES_Ala2}
	\end{figure}
	\newpage
	
	\subsection{Model sizes and $\beta_{IB}$ values}
	Using the $\beta_{IB}$ selection procedure described above, the following hyperparameters were used across all systems. The $\beta_{IB}$ values are obtained systematically; model sizes were kept minimal and increased only as needed for accurate state prediction.
	
	\begin{table}
		\centering
		\begin{tabular}{|c|c|r|}
			\hline
			\textbf{System} & $\mfocus{\beta_{IB}}$ \textbf{value} & \textbf{Model size (Encoder, Decoder)} \\ \hline
			Radial Double-Wells & 0.1   & 1024,1024        \\ \hline
			Alanine Dipeptide & 0.25   & 256,256        \\ \hline
			\Aib & 0.05   & 512,256        \\ \hline
			
		\end{tabular}
		\caption{Hyperparameters used for all benchmark systems}
		\label{SI-TABLE-PARAMS}
	\end{table}
	\newpage
	
	\section{Additional results}
	Additional results are shown below.
	
	\subsection{Projecting Alanine Dipeptide FES along $\psi$}
	Although $\phi$ was the primary coordinate used for state labeling during SPIB training, the GFlowNet samples can also be projected onto $\psi$, reproducing the FES with high accuracy.
	\begin{figure*}[h]
		\includegraphics[width=1.0\linewidth]{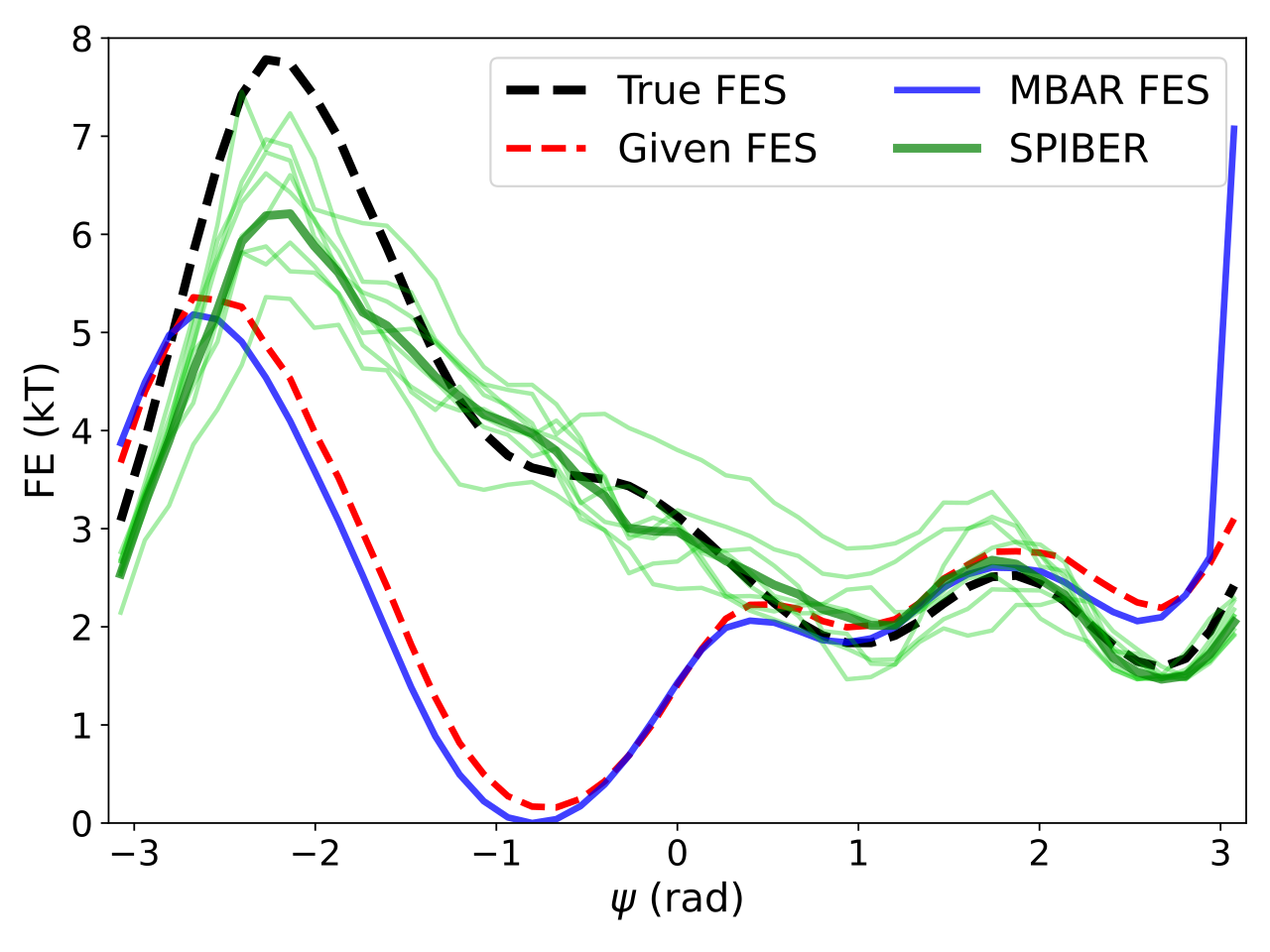}
		\caption{FES projected from \methodname~samples along $\psi$. The agreement with ground truth is strong, particularly near the minima.}
		\label{SI-PSI-ALA2}
	\end{figure*}
	\newpage
	
	\subsection{Comparing \methodname~against DHAM}
\begin{figure*}[h]
		\includegraphics[width=1.0\linewidth]{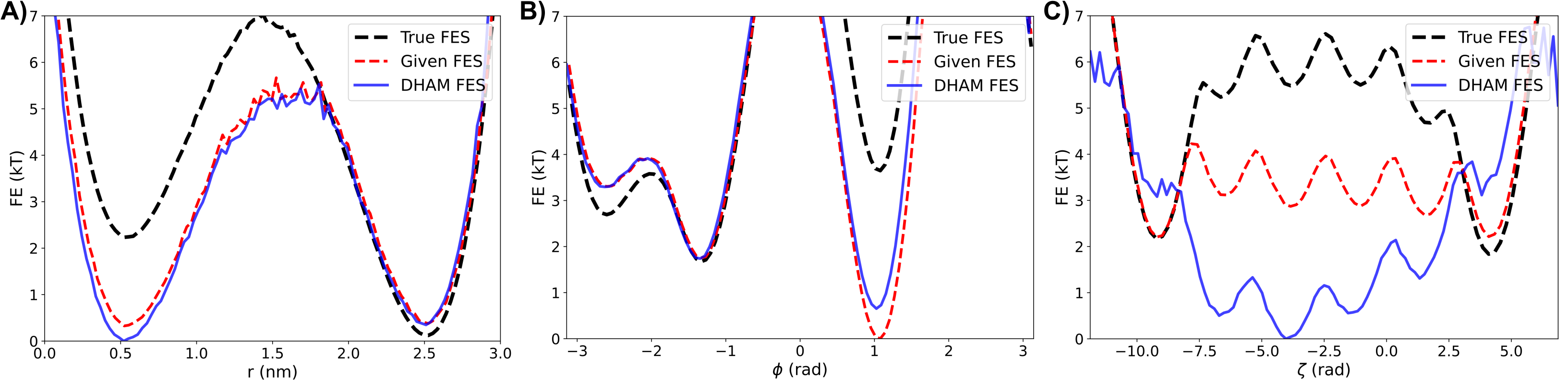}
		\caption{FES computed by \methodname~compared to DHAM for A) Radial Double-Wells B) Alanine Dipeptide and C) \Aib.}
		\label{SI-DHAM}
	\end{figure*}

	See Section IIIE of the main manuscript for a description of the DHAM procedure. As shown in Figure~\ref{SI-DHAM}, DHAM yields only marginal improvement over the raw histogram, and the results are inconsistent across systems. For \Aib (panel C), the estimate is degraded relative to the raw FES due to the oversampling of transitions inherent in our short-trajectory design. Even under more realistic sampling conditions (panels A and B), DHAM fails to meaningfully correct population imbalances.
	\newpage
	
	\subsection{Energy-based reweighting using traditional Collective Variables for Alanine Dipeptide}
	When systems of interest have major metastable states defined also by well-characterized degrees of freedom such $\phi,\psi$ for Alanine Dipeptide, using these collective variables might result in efficient energy-based reweighting. In this case, the residual entropy after specifying $\phi,\psi$ is nearly constant, and so this set forms a reasonable basis for energy-based reweighting. However, utilizing a CV that is not entropy-free can result in poor or no reweighting at all, as in the case of ($\psi,\theta$).
	\begin{figure}
		\captionsetup{skip=6pt}
		\centering
		\includegraphics[width=0.95\textwidth]{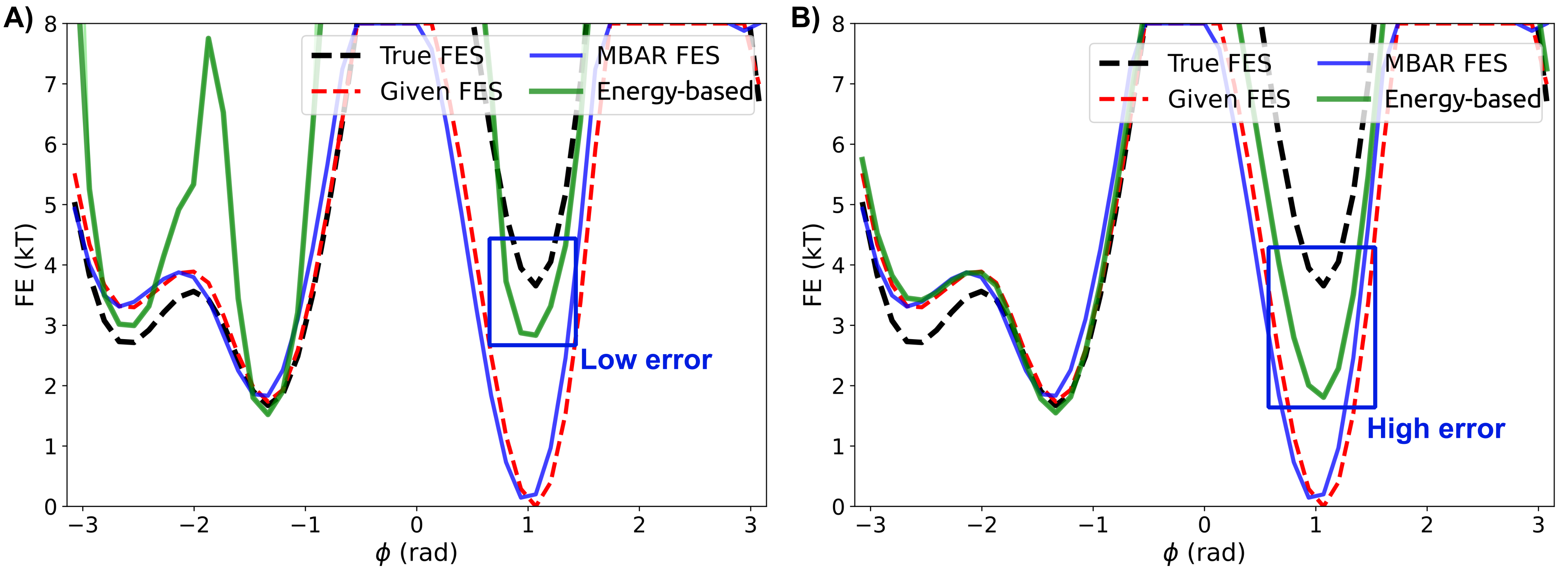}
		\caption{(A) Reweighting with $\phi,\psi$ as CVs works reasonably near metastable states, since these dihedrals capture nearly all conformational freedom and $F(z) \approx \hat{E}(z) + C$ approximately holds. (B) Using $\psi,\theta$ instead fails, because entropy depends on the unobserved $\phi$.}
		\label{SI-FES_Ala2_PhiPsi}
	\end{figure}
	\newpage
	
	\subsection{Poor performance of reweighting \Aib~with $\zeta$}
\begin{figure}
		\captionsetup{skip=6pt}
		\centering
		\includegraphics[width=0.9\textwidth]{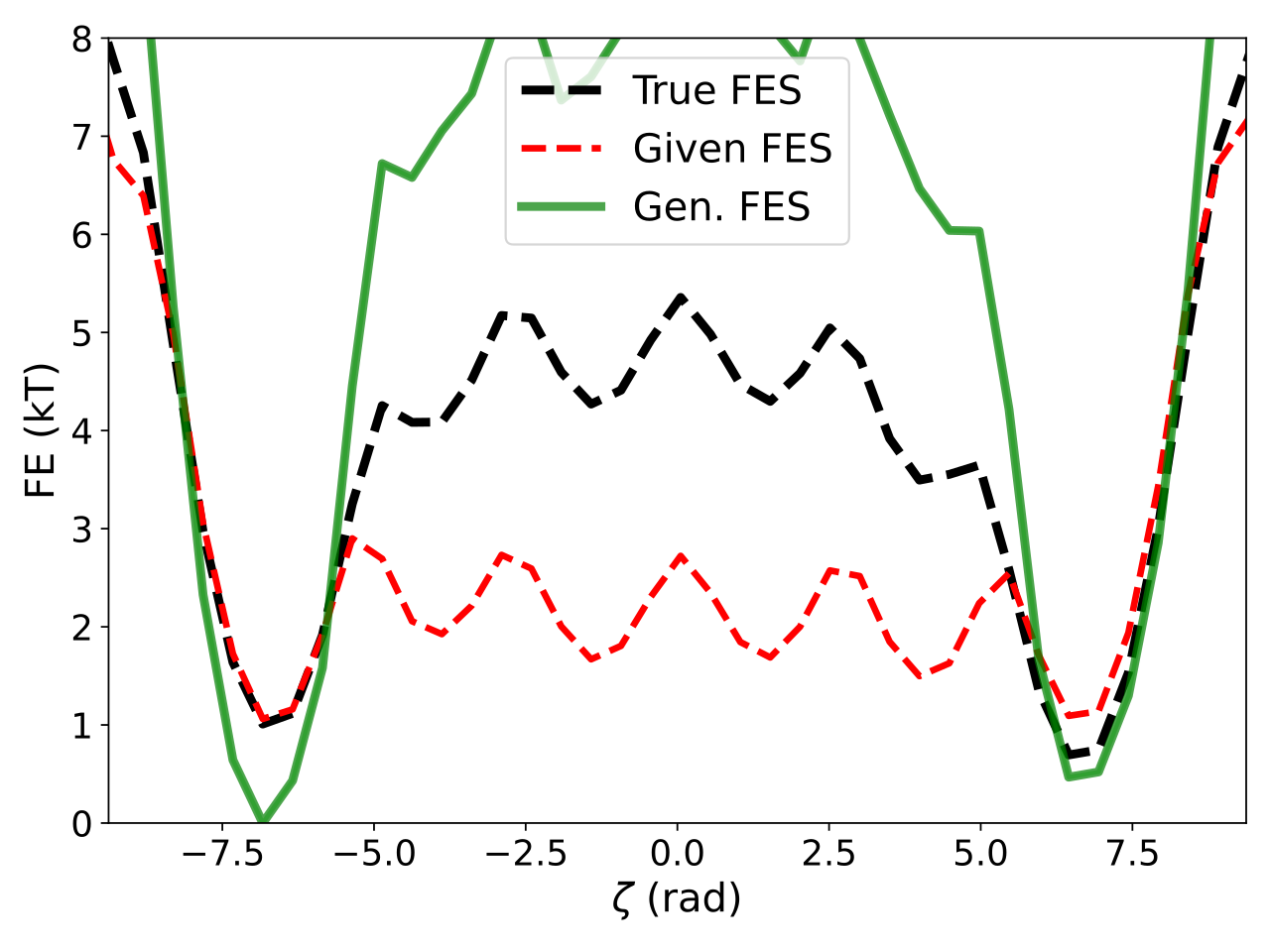}
		\caption{Reweighting with $\mfocus{\zeta}$ as a 1D CV fails, because the degeneracy of $\zeta$ leads to large, state-dependent conditional entropy.}
		\label{SI-AIB9_Results_Zeta}
	\end{figure}

	$\zeta$ is a known variable that is able to separate key metastable regions of \Aib~as shown in Figure~\ref{SI-AIB9_Results_Zeta} below (black curve). However, this CV is not entropy-free. Demonstrably, states with $\zeta \approx \pm 5$ have much lower entropy since the helix is fully formed. States with $\zeta \approx \pm 1$ have the highest entropy where all but one of dihedrals in \Aib~are unconstrained. This results in missing entropic corrections for the intermediate well, resulting in the predicted free energy being much higher than the true value. This illustrates that intuitively chosen CVs are often unsuitable for energy-based reweighting.
